\documentclass[11pt]{wlscirep}
\usepackage[utf8]{inputenc}
\usepackage[T1]{fontenc}
\usepackage[final]{pdfpages}

\title{Complexity of the COVID-19 pandemic in Maring\'a}

\author[1]{Andre S. Sunahara}
\author[1]{Arthur A. B. Pessa}

\author[2,3,4,5,6,*]{Matja{\v z} Perc} 
\author[1,$\dagger$]{Haroldo V. Ribeiro} 

\affil[1]{Departamento de F\'isica, Universidade Estadual de Maring\'a - Maring\'a, PR 87020-900, Brazil}

\affil[2]{Faculty of Natural Sciences and Mathematics, University of Maribor, Koro{\v s}ka cesta 160, 2000 Maribor, Slovenia}
\affil[3]{Department of Medical Research, China Medical University Hospital, China Medical University, Taichung, Taiwan}
\affil[4]{Alma Mater Europaea, Slovenska ulica 17, 2000 Maribor, Slovenia}
\affil[5]{Department of Physics, Kyung Hee University, 26 Kyungheedae-ro, Dongdaemun-gu, Seoul, Republic of Korea}
\affil[6]{Complexity Science Hub Vienna, Josefst{\"a}dterstra{\ss}e 39, 1080 Vienna, Austria}

\affil[*]{email: matjaz.perc@gmail.com}
\affil[$\dagger$]{email: hvr@dfi.uem.br}

\begin{abstract}
While extensive literature exists on the COVID-19 pandemic at regional and national levels, understanding its dynamics and consequences at the city level remains limited. This study investigates the pandemic in Maring\'a, a medium-sized city in Brazil's South Region, using data obtained by actively monitoring the disease from March 2020 to June 2022. Despite prompt and robust interventions, COVID-19 cases increased exponentially during the early spread of COVID-19, with a reproduction number lower than that observed during the initial outbreak in Wuhan. Our research demonstrates the remarkable impact of non-pharmaceutical interventions on both mobility and pandemic indicators, particularly during the onset and the most severe phases of the emergency. However, our results suggest that the city's measures were primarily reactive rather than proactive. Maring\'a faced six waves of cases, with the third and fourth waves being the deadliest, responsible for over two-thirds of all deaths and overwhelming the local healthcare system. Excess mortality during this period exceeded deaths attributed to COVID-19, indicating that the burdened healthcare system may have contributed to increased mortality from other causes. By the end of the fourth wave, nearly three-quarters of the city's population had received two vaccine doses, significantly decreasing deaths despite the surge caused by the Omicron variant. Finally, we compare these findings with the national context and other similarly sized cities, highlighting substantial heterogeneities in the spread and impact of the disease.
\end{abstract}

\begin{document}

\rfoot{\small\sffamily\bfseries\thepage/20}%

\flushbottom
\maketitle
\thispagestyle{empty}

\section*{Introduction}

The coronavirus disease 2019 (COVID-19) is an infectious disease caused by the novel severe acute respiratory syndrome coronavirus 2 (SARS-CoV-2)~\cite{zhu2020anovel}. The first case of COVID-19 was identified in Wuhan on 1 December 2019. Subsequently, the virus spread rapidly worldwide~\cite{delatorre2020tracking}, prompting the World Health Organization (WHO) to officially declare a pandemic on 11 March 2020~\cite{delatorre2020tracking}. In Brazil, the initial case was officially reported on 25 February 2020, in S\~ao Paulo. Nevertheless, evidence suggests that the virus had been circulating in Brazil much earlier~\cite{candido2020evolution}, with a retrospective study of serum collected from thousands of patients in the state of Esp\'irito Santo indicating the presence of COVID-19 antibodies already in mid-December 2019~\cite{stringari2021covert}. However uncertain the spread of the virus started, COVID-19 had reached virtually all Brazilian municipalities by mid-August 2020 with 70\% of them having registered at least one death~\cite{ribeiro2020city}. 

COVID-19 arrived in the largest metropolitan areas of the Brazilian Southeast, namely S\~ao Paulo and Rio de Janeiro, most likely imported from Europe and China~\cite{candido2020routes, candido2020evolution}. Subsequently, the disease rapidly disseminated through air and ground transportation to the interior regions of the country, initially affecting the North and Northeast macroregions before spreading to the Midwest and South~\cite{castro2021spatiotemporal, candido2020evolution, candido2020routes, ranzani2021characterisation}. In the North, the region with the fewest ICU beds~\cite{amib-boletim}, the public healthcare system of Manaus had already collapsed by May 2020~\cite{fiocruz_boletim}, foreshadowing the country's worst pandemic period, which extended from late 2020 to the end of 2021~\cite{boschiero2021oneyear, fiocruz_boletim}. As the pandemic progressed, the epicenter of the disease shifted repeatedly, resulting in an overall description of three major waves of infections and deaths in Brazil between March 2020 and January 2022~\cite{fiocruz_boletim}.

By the end of 2022, Brazil had recorded 36.3 million confirmed cases of COVID-19, accounting for approximately 5\% of global cases~\cite{ourworldindata_covid}. Alarmingly, the virus claimed 694 thousand lives in Brazil, representing approximately 10\% of all COVID-related deaths worldwide~\cite{ourworldindata_covid}. This mortality figure is an undeniably disproportionate toll, given that Brazil comprised less than 3\% of the world's population in 2021~\cite{ourworldindata_population}. The pandemic has had far-reaching consequences, manifesting in dramatic reductions in life expectancy among both newborns and older citizens in 2020 and 2021, particularly in the northern part of the country~\cite{castro2021reduction}.

Cities in Brazil have played a noteworthy role in managing the COVID crisis following the Supreme Court's decision against the centralization of decision-making by the federal government in late March 2020~\cite{rodrigues2021brazil, biehl2021supreme}. This decentralization was motivated not only by the vast geographic extension of the country and nonconcurrent patterns of spread~\cite{castro2021reduction, candido2020evolution} but also by political issues that hindered cooperation and coordination between different government levels~\cite{de2022quantifying}. While decentralization enables cities to swiftly and effectively develop tailored responses to the pandemic, it may also exacerbate local socioeconomic disparities in a country as diverse as Brazil. Moreover, despite the unusual aspect of this decentralization of decision-making in Brazil and the current advanced pace of the pandemic, there is still a lack of comprehensive understanding of the disease dynamics and its consequences at the city level. This sharply contrasts with abundant literature exploring the COVID-19 pandemic at the regional and national levels and mainly reflects the considerable difficulties in obtaining local data beyond traditional variables such as cases and deaths. For instance, when available, information related to the occupancy of hospital beds and administration acts tend to be publicized by cities through social media in irregular formats (e.g., newsletters and dedicated webpages) that hampered the consolidation of this type of data into a national database. In many cases, this type of information was not even recorded but only daily updated, such that keeping track of these records would require active monitoring.

This work aims to bridge this gap by comprehensively investigating the COVID-19 pandemic in Maring\'a, a medium-sized Brazilian city with over 400,000 inhabitants and a high Human Development Index (0.808, 2010 census) located in the state of Paran\'a, South Region of Brazil. To accomplish this, we manually curated a dataset by actively tracking all newsletters published by the city administration on a daily basis over 817 days between 18 March 2020 and 12 June 2022, alongside other data obtained from the Oswaldo Cruz Foundation (Fiocruz), Civil Registry, and Google. Our dataset extends beyond conventional variables such as the number of COVID-19 cases and deaths, encompassing daily hospital bed occupancy, all city administration acts that imposed and relaxed non-pharmaceutical interventions, progress in vaccine coverage, and mobility indicators. Furthermore, by meticulously analyzing all city decrees, we classify the imposed restriction into groups, and by attributing a score to each one, we propose an aggregated measure of the intensity imposed on the citizens.

Among our findings, we report that non-pharmaceutical interventions were effective in significantly reducing human mobility within the city but did not prevent cases from rising exponentially at the beginning of the pandemic, with a reproduction number slightly below estimates from the initial outbreak in Wuhan. We identify six distinct waves of COVID-19 cases in Maringá, each marked by significant disparities in both deaths and hospital bed occupancy. The third and fourth waves were the deadliest, accounting for 68\% of all deaths recorded in our dataset. These waves also overwhelmed the local healthcare system, resulting in excess mortality that cannot be solely attributed to COVID-19 deaths. Towards the end of the fourth wave, vaccination efforts had increased substantially, with three-quarters of the city's population having received two doses of the vaccine. Despite the impressive surge in cases caused by the Omicron variant, the fifth and sixth waves were characterized by a more manageable situation with substantially fewer deaths. Our research shows that non-pharmaceutical interventions had a remarkable impact on both mobility and pandemic indicators, and these effects were most pronounced during the onset of the pandemic and the most severe phase of the pandemic. However, our results also suggest that the city's measures were more reactive than proactive in response to the pandemic situation. Furthermore, we compare the pandemic evolution in Maring\'a with other similarly sized Brazilian cities and the national picture, revealing significant heterogeneity in the spread and impact of the virus.

In what follows, we provide a detailed investigation of the onset of the pandemic and subsequent waves of COVID-19 cases in the city. We then present a meticulous analysis of the city's responses to the ongoing pandemic, including the aggregated measure of the intensity of non-pharmaceutical interventions and its association with pandemic and mobility variables. Finally, we compare some of our findings with the national trends and other cities of similar size. We conclude by summarizing and discussing our results. The Methods section of this article provides a detailed description of the processes used to build our dataset, as well as the models we employ to estimate epidemiological parameters.

\section*{Results}

\subsection*{Exponential growth and the pandemic's beginning} \label{sec:beginning}

We start by analyzing data on the number of cases during the initial spread of COVID-19 in Maring\'a. This data was gathered from daily reports issued by the city administration and covers a period of 817 days between 18 March 2020 and 12 June 2022 (see Methods for details). The first case of COVID-19 in the city was confirmed on 18 March 2020, almost one month after the disease had first appeared in Brazil. This first case involved a 45-year-old woman from Spain who arrived in the city on 11 March 2022. Maring\'a was the 58th out of 5,570 Brazilian cities to report a COVID-19 case. Following the confirmation of the first case, the city administration declared a state of emergency and imposed restrictions, including the closure of schools and nonessential businesses. The inter-municipal transportation system was suspended and a curfew (stay-at-home order) was put in place from 21h to 5h to limit public movement. These measures were in line with international responses to the disease and reflected the situation in China and Italy~\cite{ren2020pandemic}. During this period, non-pharmaceutical interventions, such as social distancing, case isolation, personal hygiene, and shielding, were the only public policies available to contain the early spread of COVID-19~\cite{markel2007nonpharmaceutical, ferguson2020report}, mainly due to the limited understanding of transmission mechanisms at that time~\cite{rahman20201thetransmission}.

\begin{figure*}[!ht]
\centering
\includegraphics[width=0.95\textwidth]{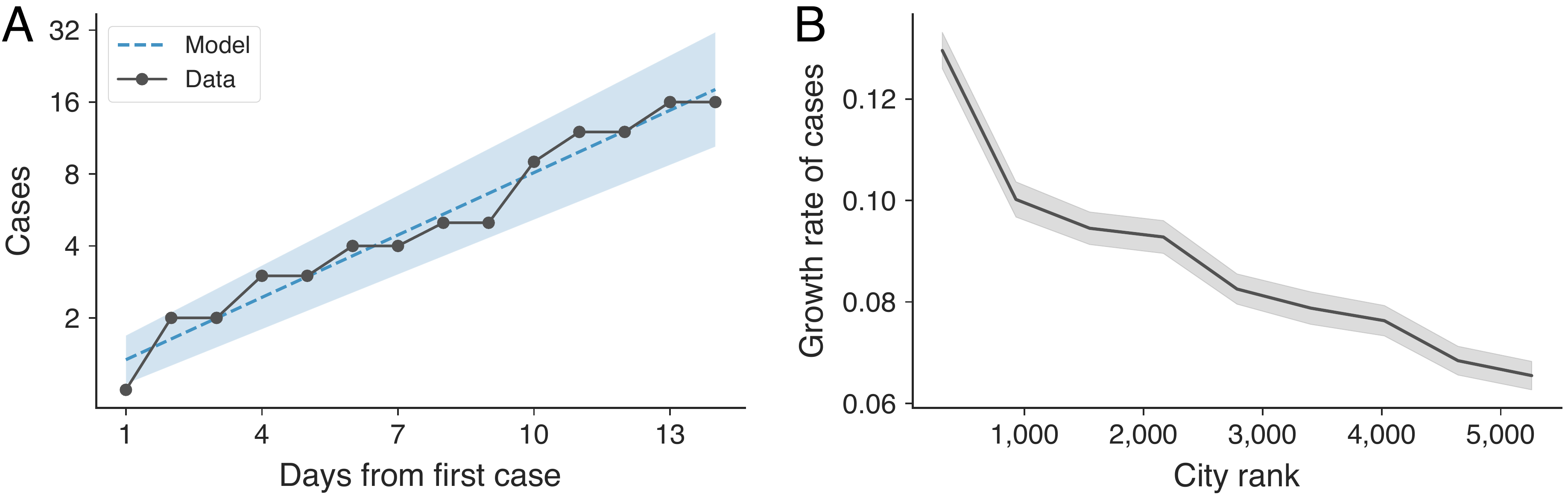}
\caption{Exponential growth in the early number of COVID-19 cases. (A) Cumulative number of cases of COVID-19 in Maring\'a during the initial 14 days since the city's first reported case (black markers). The dashed line represents the adjusted exponential model (Eq.~\ref{eq:exp}) and the shaded region indicates one standard deviation band of the model. (B) Association between the estimated initial exponential growth rate of cases for the initial 14 days since the first reported case and city population rank (the lower the rank, the higher the population). The shaded region indicates the standard deviation of the mean. On average, larger cities exhibit higher exponential growth rates of cases, while smaller towns display lower rates.}
\label{fig1}
\end{figure*}

Despite timely efforts to contain the disease spread, we find an exponential increase in the number of COVID-19 cases during the first two weeks since the first case. Figure~\ref{fig1}{A} shows the cumulative number of cases in comparison with the exponential model with growth rate $r=0.2\pm0.1$ (see Eq.~\ref{eq:exp} in Methods) adjusted from data. This means that the number of cases would double every $\approx 3.5$~days if this exponential growth persisted. Past research has also verified that this initial growth rate increases with city size~\cite{ribeiro2020city, stier2021early}. Figure~\ref{fig1}{B} illustrates the relationship between the growth rate estimated from the initial fourteen days after the first case for all Brazilian cities and their population rank (the lower the rank, the higher the population). As also reported in Ref.~\cite{ribeiro2020city}, we observe a decreasing trend in the growth rate with city rank. For instance, Maring\'a was the 55th largest population and displayed a growth rate of $r=0.20$, while S\~ao Paulo -- the most populous Brazilian city -- exhibited a 20\% higher growth rate ($r=0.24$). It is also worth noticing that although the average growth rate monotonically diminishes with city rank, there is significant variability between cities of similar size, likely due to factors such as reporting delays, under-reporting, differences in social distancing policies, and socioeconomic characteristics~\cite{fiocruz_boletim, boschiero2021oneyear, sutton2022population}.

Another key epidemiological variable related to the initial growth rate is the basic reproduction number or the reproductive number $\mathcal{R}_0$. The value of $\mathcal{R}_0$ represents the expected number of people one sick individual will infect. For $\mathcal{R}_0 < 1$, the number of cases will decrease exponentially, while $\mathcal{R}_0 > 1$ indicates an exponential rise in disease transmission with amplification of cases~\cite{stier2021early}. Mathematically, $\mathcal{R}_0$ can be defined as $\mathcal{R}_0=1+r/\gamma$, where $r$ is the initial growth rate and $1/\gamma$ is the infectious period (see Methods for further details). The infectious period represents the time interval during which a sick individual can transmit the disease to susceptible individuals. For COVID-19, the estimated infectious period is $4{.}8$ days~\cite{nishiura2020serial}. Using this value and the growth rate of $r = 0.20$, we calculate the reproductive number of $\mathcal{R}_0 = 1.96$ for Maring\'a in the first fourteen days after the first case. By varying the length of the initial exponential period from seven to thirty days and fitting exponential functions to the number of cases, we find point estimates for $\mathcal{R}_0$ ranging from $1.61$ to $2.10$. Although these values are below the estimates obtained from the initial outbreak in Wuhan, which ranged from $2.2$ to $6.5$~\cite{liu2020thereproductive, wu2020nowcasting}, they are still quite close to the lower bound.

After the initial phase of an epidemic, it is not possible to estimate $\mathcal{R}_0$ as the population is continually exposed to the virus and many individuals acquire immunity through infection. However, an alternative, yet related, measure that has been extensively used to monitor the pandemic's evolution and the impacts of public policies is the instantaneous reproduction number, denoted as $\mathcal{R}(t)$ (see Methods for further details). This measure represents the average number of infections caused by an infected individual at time $t$ and has been proven highly sensitive to changes in the transmission rate~\cite{fraser2007estimating}, as we shall discuss in the subsequent section.

\subsection*{Temporal heterogeneity and waves of cases}

\begin{figure*}[!ht]
\centering
\includegraphics[width=0.95\textwidth]{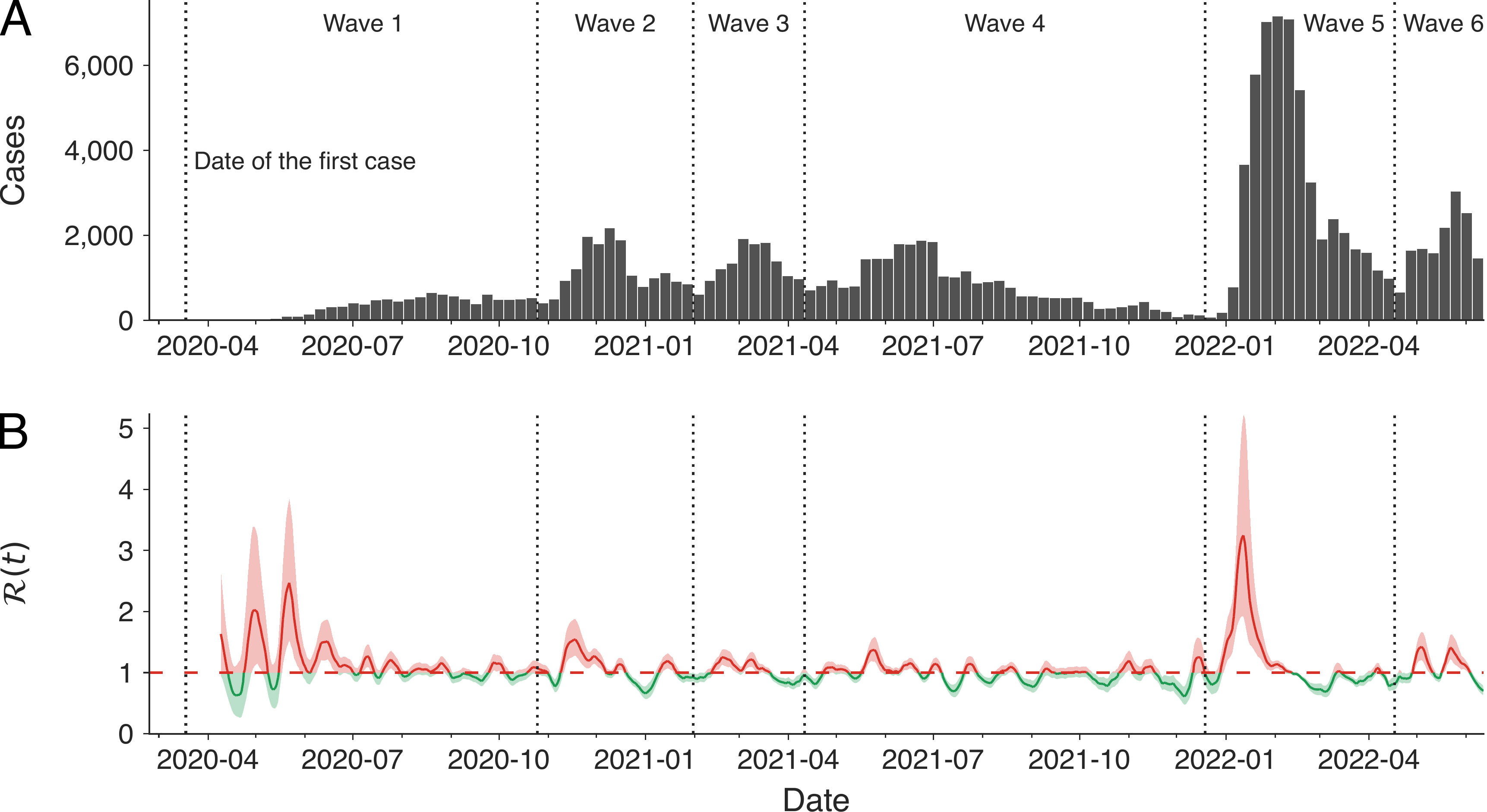}
\caption{Evolution of COVID-19 cases and instantaneous reproduction number in the city. (A) The bars show the weekly numbers of confirmed cases between 18 March 2020 and 12 June 2022. (B) The continuous line depicts the dynamics of the instantaneous reproduction number ($\mathcal{R}(t)$, as described in Methods) from 13 April 2020 to 12 June 2022. Shaded regions represent the 95\% confidence intervals, and the dashed horizontal line indicates the epidemic threshold $\mathcal{R}(t) = 1$. The curve corresponds to a 7-day moving average, and the color code indicates reproduction numbers above (red) or below (green) the epidemic threshold. In both panels, vertical dashed lines separate the six identified waves of cases, which represent periods of successive rise and fall in the weekly number of cases lasting more than four weeks.}
\label{fig2}
\end{figure*}

The complete evolution of the weekly number of COVID-19 cases in Maring\'a, from March 2020 and June 2022, is illustrated in Figure~\ref{fig2}{A}. We observe that the number of cases displayed a series of epidemic waves. Although there is no established working definition or consensus on the minimum requirements to define an epidemic wave~\cite{zhang2021second, ayala2021identification}, we consider a wave as a period of successive rise and fall in the weekly number of cases lasting more than four weeks. With this criterion, we identify six waves, as indicated by vertical dashed lines in Figure~\ref{fig2}{A}. Applying the same definition, we find that four waves occurred nationally during the same period (Figure~S1{A}), highlighting that the nonsynchronous spread dynamics in each city~\cite{castro2021reduction, candido2020evolution} generate an aggregated pattern that not necessarily represents the local situation. We also estimate the evolution of the instantaneous reproduction number $\mathcal{R}(t)$ throughout the entire period covered by our data (see Methods), as depicted in Figure~\ref{fig2}{B}. In retrospect, the dynamics of $\mathcal{R}(t)$ provided a reliable indicator of changes in the number of cases. Typically, this indicator was below one at the end of each wave and exhibited a peak a few weeks before the number of cases reached a local maximum.

\begin{figure*}[!ht]
\centering
\includegraphics[width=0.85\textwidth]{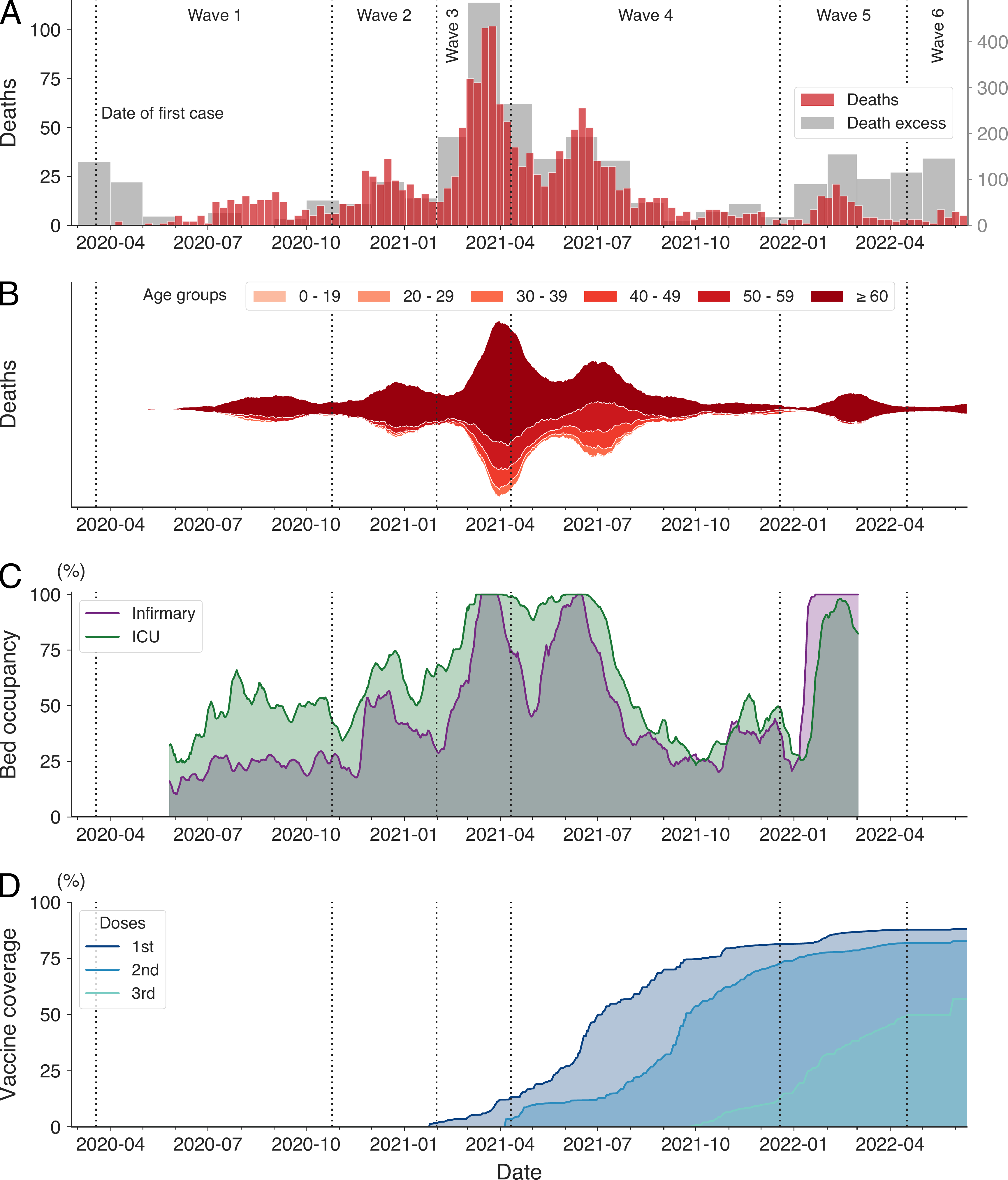}
\caption{Evolution of COVID-19 mortality, hospital bed occupancy, and vaccine coverage in the city. (A) The red bars depict the weekly COVID-19 death toll between 18 March 2020 and 12 June 2022. The gray bars in the background represent monthly excess mortality. This metric refers to the disparity between the number of deaths recorded in a particular month and the number reported in the same month of 2019 (pre-COVID-19 emergence). (B) The stacked area chart shows the proportion of COVID-19 deaths classified into six age groups as indicated by the color code. The curves correspond to monthly moving averages. (C) Daily occupancy percentage of infirmary (purple curve) and intensive care unit (ICU, green curve) beds, exclusive to COVID patients, between 22 May 2020 and 2 March 2022. The curves are 7-day moving averages. (D) Evolution of the COVID-19 vaccine coverage in the city grouped by number of doses administered (as indicated by the color code) between 19 January 2021 and 12 June 2022. In all panels, vertical dashed lines delineate the six identified waves of COVID-19 cases.}
\label{fig3}
\end{figure*}

In addition to analyzing the number of cases, we investigate the complete evolution of the number of deaths in the city. Figure~\ref{fig3}{A} displays the weekly number of deaths as well as the excess mortality with monthly resolution. Additionally, Figure~\ref{fig3}{B} presents the weekly number of deaths categorized into six age groups (0-19, 20-29, 30-39, 40-49, 50-59, and with 60 years old or older). These figures reveal that the waves of cases roughly correspond to the waves observed in the number of deaths. In total, 1,847 people died from COVID-19 during the period covered by our data. However, there is substantial heterogeneity in the distribution of deaths over time and among different age groups. We have also closely monitored the occupancy of infirmary and intensive care unit (ICU) beds exclusive to COVID patients, as well as the percentage of the city's population immunized with one, two, and three doses of COVID-19 vaccines. Figure~\ref{fig3}{C} shows the daily percentage of hospital bed occupancy for the two categories, while Figure~\ref{fig3}{D} depicts the evolution of vaccine coverage. Data for hospital bed occupancy started to be released by the city administrations on 22 May 2020 (65 days after the first case) and were not reported beyond 2 March 2022.

Combining the results shown in Figures~\ref{fig2} and \ref{fig3} provides a detailed examination of the different pandemic stages that took place in the city. Starting from the first wave, which extended from late March to 25 October 2020, we find that the evolution of the number of cases closely resembles the first wave of cases in Brazil~\cite{fiocruz_boletim} (Figure~S1{A} for comparison with the national number of cases). During this first wave, there were 315 cases per week, and a peak of 657 weekly cases was recorded by the end of August 2020. Over the same period, there were 6 weekly deaths, with a peak of 17 occurring in the week ending on 6 September, right after the peak in the number of cases. About 80\% of all fatalities in the first wave corresponded to citizens 60 years old or older. This much-imbalanced death toll was a feature of the COVID-19 pandemic~\cite{ranzani2021characterisation, souza2021analysis} and informed future policies, particularly regarding vaccines~\cite{fiocruz_boletim}. Men were also more commonly affected by COVID and amounted to 56\% of deaths during this first wave of cases in Maring\'a (Table~S1). The occupancy of hospital beds remained at low levels during the first wave, but we observe a high excess mortality at the beginning of the first wave (139 only in March 2020). While it is difficult to attribute a cause to this behavior, underreporting of COVID deaths, earlier circulation of the virus in the city (as was also observed at the country level~\cite{candido2020evolution, stringari2021covert}) and avoidance in seeking medical care due to the initial shock and confusion caused by the pandemic are possible explanations.

After the initial wave of COVID-19 infections, 2020 ended with a surge of cases that marked the onset of a second wave from 26 October 2020 to 31 January 2021. This wave highlighted the virus's ever-mutating characteristics, as the Gamma variant originated in Manaus~\cite{faria2021genomics} became dominant in all Brazilian states during the country's worst period of the health crisis~\cite{fiocruz_boletim, fiocruz-genomas}. Despite the reproduction number during this wave being significantly lower than that observed in the first wave (maximum of 1.54 versus 2.47), weekly cases and deaths were substantially higher, with an average of 1,191 cases and 18 deaths per week. Additionally, hospital bed occupancy was higher than during the previous wave, although it did not reach full capacity. Senior citizens were again disproportionately affected by the disease, accounting for 86\% of deaths in this wave (Table~S1). Consequently, in accordance with national guidelines~\cite{fiocruz_boletim}, the city administration developed an immunization plan that prioritized senior citizens over 60 years of age and health professionals, followed by people with comorbidities or who were socially vulnerable, and progressively reaching the younger population. A health professional was the first person to receive a vaccine in the city on 19 January 2021, only two days after the first dose was administered in the country~\cite{fiocruz_boletim}.

In February 2021, cases began to increase again, signaling the onset of a short third wave lasting from 1 February to 11 April 2021. While the number of cases was similar to that of the second wave, the number of deaths rose substantially once more. The third wave averaged 57 deaths per week, three times more than the second wave and almost ten times more than the first wave. This wave recorded the highest weekly number of deaths in the city (which also coincided with the deadliest moment of the pandemic in Brazil~\cite{fiocruz_boletim, ourworldindata_covid}) and alone accounted for 31\% of all deaths covered by our data, with a peak of 102 deaths occurring in the epidemiological week ending on 28 March 2021 (almost 6\% of all COVID deaths). Impressively, the third wave more than doubled the accumulated number of deaths from 434 to 1,006, with 572 deaths in contrast to 184 and 250 deaths in the first and second waves, respectively. This sharp increase in mortality also affected the case fatality rate, which rose from 1.62\% in the first two waves to 4.37\%. The third wave also marked the beginning of an extended period of high hospital bed occupancy, with the health system reaching and remaining at full capacity for ICU beds for almost six weeks. During this wave and the subsequent one, we observe significant variations in the total number of exclusive infirmary and ICU beds (public and private) available to accommodate the increasing number of people infected with COVID-19 (Figure~S2). March 2021 recorded the highest excess deaths with 486 more deaths than the same period of 2019. Even projecting a month with 102 weekly deaths, the weekly mortality peak, we could not fully explain this high excess mortality, suggesting that the overloaded healthcare system contributed to the rise of mortality due to other causes. With immunization progressing slowly and focusing on senior citizens, we began to observe a change in the profile of COVID deaths. Citizens over 60 years old accounted for 70\% of all deaths during this third wave, a reduction from fractions larger than $\approx80$\% during the previous two waves (Table~S1).

The fourth wave of cases started on 12 April 2021 and lasted until 19 December 2021. Initially, this wave was characterized by a decrease in the number of deaths and hospital bed occupancy, which was short-lived, as both metrics soon started to rise again, culminating in another peak towards the end of June 2021. Hospital bed occupancy once again reached full capacity but remained at this critical level for a shorter duration than during the third wave. Excess mortality was significantly lower during the fourth wave, with almost all the excess deaths attributable to COVID-19. Notably, the fourth wave was also marked by an impressive increase in the number of vaccinated individuals. By the end of 2021, 81\% of the population had received one vaccine dose, 74\% had received two doses (or one dose of the one-shot Jansen vaccine), and 15\% had received three doses (mostly seniors). These high percentages are characteristic of Brazilian vaccination campaigns~\cite{fiocruz_boletim} and comparable to those of developed countries~\cite{ourworldindata_covid}. The fourth wave was the longest and accumulated the highest number of deaths, accounting for 686 deaths (37\% of all deaths) during its nine-month span. Following the changes in the profile of COVID deaths that began with the third wave, 58\% of the deaths during the fourth wave were among senior citizens, the lowest percentage of all six waves (Table~S1). The increase in deaths among younger age groups can likely be attributed to the resumption of economic activities and the initial focus of the immunization plan on older citizens~\cite{fiocruz_boletim, scholey2022life}. Gender differences in the death profile were also accentuated during the fourth wave, with men accounting for 62\% of all COVID deaths. Following the peak at the end of June 2021, hospital bed occupancy decreased significantly, reaching levels similar to those observed at the beginning of the pandemic by November 2021.

Combined, the third and fourth waves accumulated 41,820 new COVID-19 cases and 1,258 deaths, 31\% of all cases, and 68\% of all deaths recorded in our database. These two waves were dominated by the Gamma and Delta variants, which are known to increase the risk of hospitalization and death~\cite{fisman2021evaluation}. Alongside the overburdened health system~\cite{fiocruz_boletim}, this likely played a critical role in creating the deadly scenario witnessed during this period. Furthermore, we find a strong correlation of 0.86 (Spearman correlation, $p$-value~$<0.001$) between the monthly number of deaths and the monthly excess deaths during the period of the third and fourth waves. This correlation strongly suggests that COVID-19 was the major cause of excess mortality during these waves.

The year 2022 brought the fifth wave of cases and the new Omicron variant~\cite{liu2022striking}. This wave was distinguished by the largest surge in cases, with 52,721 cases reported in just four months, a new average high of 3,077 cases per week, and a peak of 7,163 cases at the beginning of February 2022. This peak represents a tenfold increase from the first wave and 328\% more cases than the previous highest peak observed during the second wave. We notice that the instantaneous reproduction number anticipated this sudden rise in transmission rate, reaching a peak value of $\mathcal{R}(t) = 3.2$ on January 13, 2022 -- three weeks before the peak in case numbers. During the peak of cases, we also observe a sharp increase in hospital bed occupancy but only infirmary beds reached full capacity during the fifth wave. The number of infirmary and ICU beds exclusively dedicated to COVID was also much smaller during this wave (Figure~S2). It is worth mentioning that information on bed occupancy ceased to be released by the city administration on 2 March 2022. Mortality was significantly lower with 122 deaths in total and 7 deaths per week on average during this wave. These numbers yielded a fatality rate of 0.23\% which in turn is much smaller than what was observed in the previous four waves. This much more manageable situation reflects the characteristics of higher transmissibility but lower risk associated with the Omicron variant~\cite{liu2022striking} and high vaccination levels in the city, which stabilized at 88\% and 82\% for the first and second doses and reached 50\% for the third dose. The fifth wave also witnessed a reversal of the trend in the profile of COVID-19 deaths, with a sharp increase in the number of senior citizens deaths (94\% of all deaths during this wave, Table~S1).

The sixth and last wave of cases documented by our dataset started on 18 April 2022 and was recorded up to 12 June 2022, when the city administration discontinued its daily reports on the COVID-19 pandemic. This wave witnessed the second-largest peak in cases, with 3,045 cases occurring in the last week of May 2022. Despite the high incidence of cases and consistently with the trends observed during the fifth wave, the number of fatalities was substantially lower during this wave than in the previous ones. A total of 33 deaths were recorded during this wave, averaging 4 deaths per week. The case fatality rate was 0.22\%, which is marginally lower than the rate observed during the previous wave. As previously mentioned, data on hospital bed occupancy by COVID patients ceased on 2 March 2022. Nonetheless, information on general bed occupancy was accessible until the city administration's last daily report. This data reveals a peak in ICU bed occupancy around the apex of cases during this wave (Figure~S3), indicating that, despite the low mortality, the sixth wave also significantly burdened the healthcare system. Moreover, we observe a high excess mortality rate of around 100 people per month starting from the fifth wave on January 2022 and extending to our most recent records. In total, 153 deaths from COVID-19 occurred during the fifth and sixth waves, which were both dominated by the Omicron variant. However, the total excess mortality amounted to 630 individuals. Even after accounting for the 133 deaths due to pneumonia during the same period, there remains an unexplained excess mortality of 344 individuals. Similar to the excess mortality observed in the pandemic beginning, attributing a cause for this excess mortality in the last two waves remains challenging. Still, possible reasons for this may include underreporting of COVID deaths, likely due to the explosive number of cases observed during these waves, and deaths indirectly related to COVID-19 and its aftereffects.

\subsection*{City responses to the ongoing pandemic}

We meticulously analyze and classify all 90 decrees issued by city administration imposing and relaxing control measures in response to the ongoing pandemic. These interventions were put in place to regulate various aspects of daily life and mitigate the spread of the virus. We classify the restrictions imposed by these decrees into the following groups: curfews (five types based on the starting and ending hours), prohibitions of public and private events (three categories depending on the number of attendees), prohibitions of religious gatherings (two categories based on the number of attendees), suspension of in-person classes (for schools and universities), prohibitions of access to bar and restaurant areas (one category each), prohibitions of access to shopping malls and most other nonessential businesses (one category each), prohibitions of access to entertainment and fitness venues (five categories: cinemas, nightclubs, theaters, clubs and associations, and gyms), and prohibitions of using outdoor spaces (two categories: open public areas and parks). 

\begin{figure*}[!ht]
\centering
\includegraphics[width=0.82\textwidth]{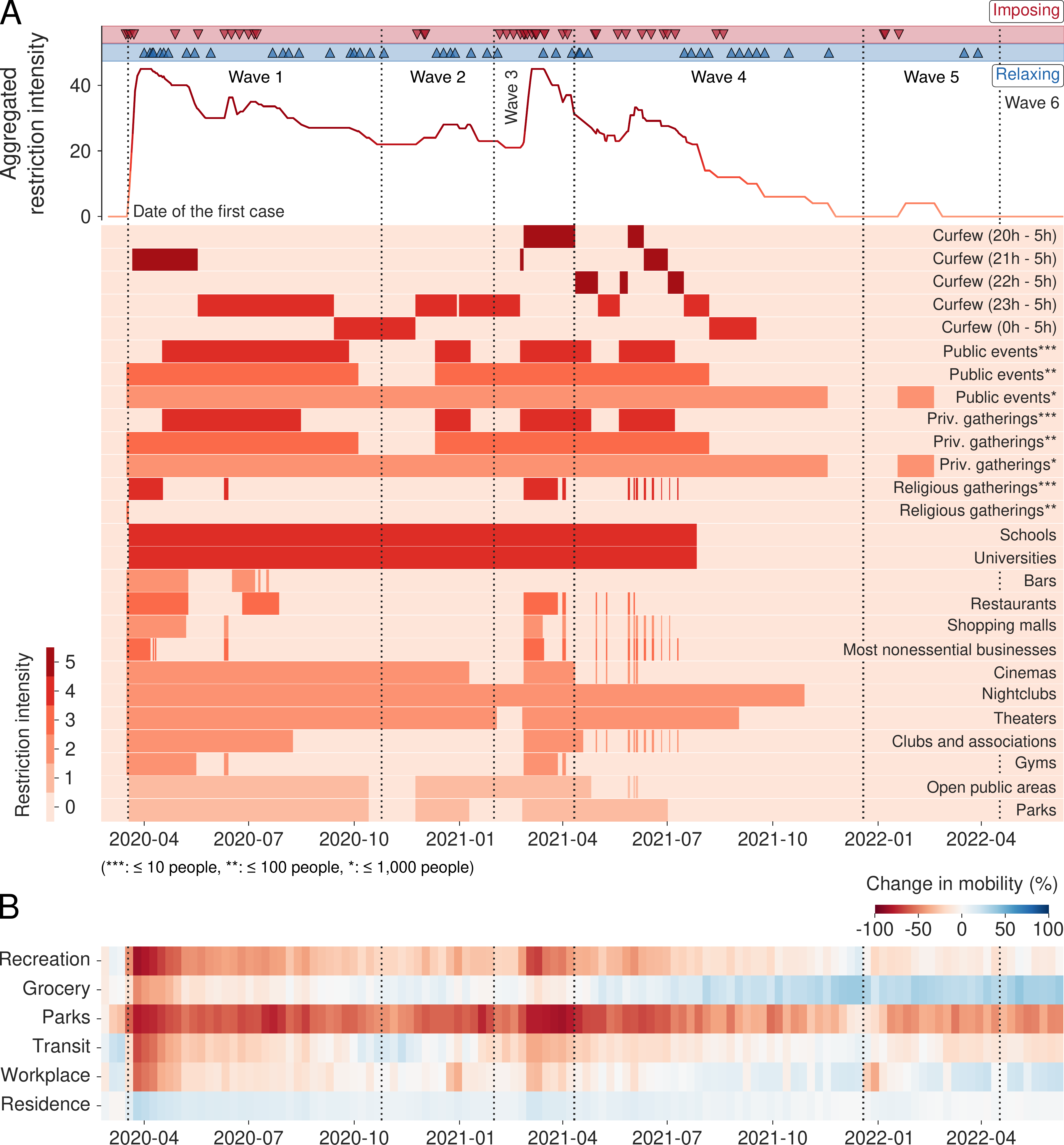}
\caption{Timeline of non-pharmaceutical interventions implemented by the city administration in response to the pandemic and changes in human mobility. (A) Temporal heatmap the activation and deactivation of each restriction category between 18 March 2020 and 12 June 2022. Each line in the heatmap corresponds to a restriction category (annotated on the right), and the color code indicates the restriction intensity (ranging from 1 to 5 in unitary steps). The curve at the top of this panel depicts the evolution of the aggregated measure of the intensity of all interventions imposed on the citizens. This indicator ranges from 0 (no restrictions in place) to 45 (all possible restrictions active) and corresponds to the sum of the scores of each restriction category. This curve is also a 7-day moving average. The triangles above the curve indicate the dates of the 90 decrees issued by city administration imposing (red) and relaxing (blue) control measures. (B) Temporal heatmap illustrating the changes in mobility related to Google users' visiting patterns to places categorized into six groups (recreation, grocery, parks, transit, workplace, and residence) compared to baselines estimated using pre-pandemic levels. Blue shades indicate an increase in the visitation to a place category, while red shades indicate a reduction. In both panels, vertical dashed lines delineate the six identified waves of COVID-19 cases.
}
\label{fig4}
\end{figure*}

Additionally, we attribute a score ranging from 1 to 5 to each of these restrictions to quantify their impact on people's lives. Higher scores indicate more stringent interventions such as curfews (rated 4 or 5 depending on their duration), while lower scores represent milder measures such as prohibition of accessing public areas and parks (rated as 1). Figure~\ref{fig4}A presents a timeline indicating when each restriction was active or not, and the color code indicates its corresponding score. Using this timeline, we create an aggregated measure of the intensity of all interventions imposed on the citizens, which corresponds to the weighted sum of the scores of each restriction. This aggregated indicator ranges from 0 (no restrictions) to 45 when all restrictions are in place. It is worth noticing that measures imposing restrictions based on thresholds in the number of people contribute to this aggregated indicator only once, meaning that only the most restrictive active measure is considered. The top panel of Figure~\ref{fig4}A also depicts the evolution of this indicator that quantifies the overall intensity of the control measures.

We also collect data from Google's community mobility reports spanning the entire period during which the city administration released its daily reports on the COVID-19 pandemic (from 8 March 2020 to 12 June 2022). As detailed in the Methods section, these data correspond to a mobility measure quantifying the change in users visiting patterns compared to baselines estimated before the COVID-19 pandemic emerged. These visiting places are further categorized into six groups: recreation, grocery, parks, transit, workplace, and residence. Figure~\ref{fig4}B depicts a temporal heatmap of the change in mobility for each category, where it is already possible to verify that the control measures significantly affected citizen mobility.

The control measures imposed using decrees were guided by a local health committee which in turn developed and regularly updated a risk matrix. This matrix considered the positivity rate of COVID tests and occupancy of ICU beds to classify the pandemic situation into six categories: low, moderate, high, very high, and extreme risk. Based on this information and in response to measures proposed by the State Government, the committee recommended control measures and public policies to be adopted by the city administration to reduce the transmission of COVID-19 in the city. It is worth remembering that cities played a prominent and atypical role during the COVID-19 crisis in Brazil due to the lack of national coordination, which was delegated to local governments by a decision of the country's Supreme Court~\cite{rodrigues2021brazil, biehl2021supreme}.

During the first wave, we observe that the city administration experimented with different control measures by issuing 31 decrees that regulated various aspects of daily life. Indeed, we note a peak in the aggregated restriction intensity only a few days after the first case was reported. These initial strict measures were motivated by all the uncertainty surrounding the characteristics and transmission mechanisms of the new disease during the early days of the pandemic. Combined with these concerns, the city's measures drastically decreased visitation to all city places and increased the permanence at home (Figure~\ref{fig4}B). However, these high levels of restrictions were not sustained as the first wave progressed, and several control measures were progressively relaxed, reaching a local minimum at the beginning of the second wave. 

Restrictions relaxed during the first wave (such as prohibitions of using outdoor spaces and events with more than ten people) were reimposed during the second wave around the peak of cases and deaths that occurred at the end of 2021. With the improvement of pandemic indicators, restrictions started to be relaxed again, reaching another local minimum during the first third of the third wave. As discussed previously discussed, the third wave was the deadliest, and the city administration issued 19 decrees during this period. Some of these decrees were in response to measures proposed by the State Government of Paran\'a, which took a much more active role during this critical moment of the pandemic (Figures~S4-S8). Similarly to what had previously occurred in the first months of the pandemic, city decrees once again heavily regulated the working hours of most types of businesses and closed many nonessential ones. From 27 February to 4 April 2021, the administration also imposed the strictest curfew until then, which, unlike previous curfews, started earlier. Public and private gatherings were restricted to a maximum of 10 people, and religious celebrations were prohibited during this wave. These measures, combined with concerns about the national situation as Brazil reached a seven-day moving average of 3,000 deaths by April 2021~\cite{ourworldindata_covid, fiocruz_boletim}, were effective in reducing mobility in the city, similarly to what was observed during the beginning of the pandemic (Figure~\ref{fig4}B).

After the most severe phase of the pandemic in the city in March 2021, some restrictions began to ease, resulting in a local minimum of aggregated restriction intensity during the early months of the fourth wave. However, hospital bed occupancy remained high throughout this period, and only restrictions on business establishments were initially relaxed during the third wave. Subsequently, there was a gradual reduction in the curfew's initial hours (from 20h to 22h and then to 23h), as well as an increase in the maximum number of individuals allowed in public and private gatherings (from 10 to 100 people). Nevertheless, as the pandemic worsened, reaching its second-deadliest moment in mid-June 2021, stricter restrictions were once again imposed, including the implementation of the strictest curfew. These more stringent measures remained in effect for a shorter period and were mostly lifted after the second half of the fourth wave. Notwithstanding, certain restrictions remained in place, such as a curfew from midnight to 5h that lasted until mid-September 2021, a prohibition on attending theaters (lifted in early September 2021), nightclubs (lifted in November 2021), and general events with more than one thousand people (lifted in mid-November 2021). Notably, it was also during this period that the city administration decreed a return to in-person classes, which began on 28 July 2021, lifting one of the longest restrictions that had been continuously active for almost 16 months.

By the end of the fourth wave, 20 months after the first case, citizens were not subject to any restrictions. This situation changed only after a surge in cases during the fifth wave. However, there was only a measure prohibiting events with more than one thousand people between 19 January 2022 and 19 February 2022. On 17 March 2022, the city administration lifted the obligatory use of masks indoors. Subsequently, on 29 March 2022, this lifting was extended to outdoor locations. During the sixth wave, the city administration did not issue any decrees, despite the surge in cases and hospital bed occupancy. Interestingly, mobility data for both the recreation and parks categories did not return to pre-pandemic levels, suggesting that visitation to public parks and some nonessential businesses remained significantly affected, even two years after the onset of the pandemic.

\begin{figure}[!ht]
\centering
\includegraphics[width=0.9\textwidth]{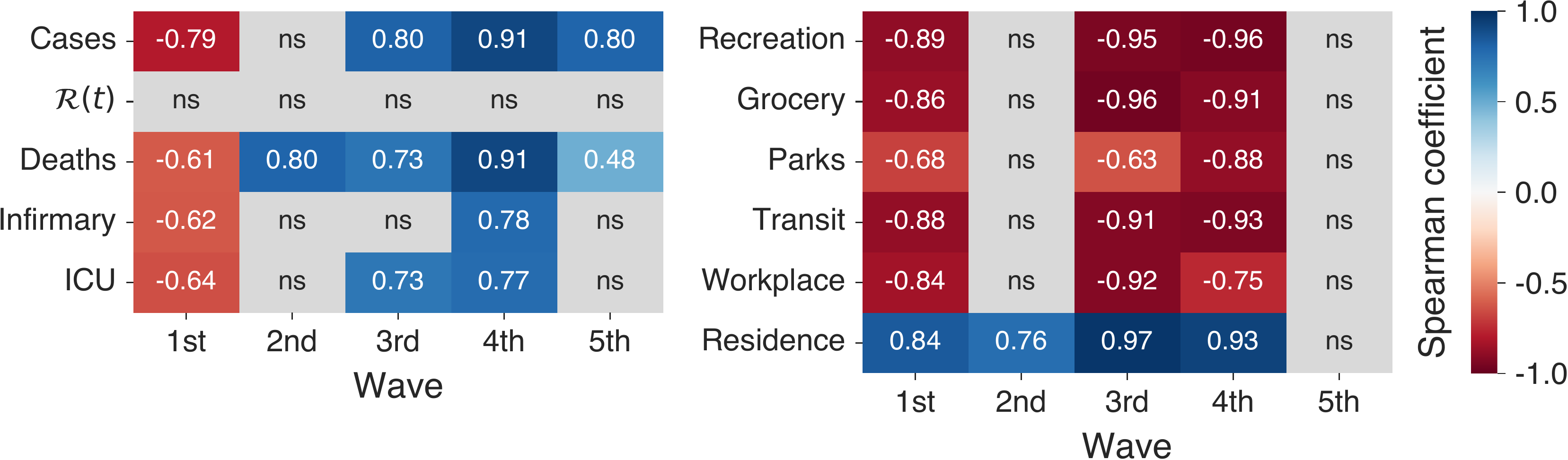}
\caption{Quantifying the association between the aggregated restriction intensity and each pandemic and mobility variable. The heatmaps present the Spearman correlation coefficient between the time series of aggregated restriction intensity and the time series related to pandemic (left panel) or mobility (right panel) variables broken down into the identified waves of COVID-19 cases. In both panels, each row refers to a pandemic or mobility variable, while each column represents a wave. The color scheme denotes the correlation value, with blue shades representing statistically significant positive correlations and red shades representing statistically significant negative correlations. Grey cells indicate nonsignificant correlations ($p$-value~$> 0.05$).
}
\label{fig5}
\end{figure}

In addition to the detailed description provided previously, we quantify the impact of the restriction measures by estimating the Spearman correlation coefficient between the aggregated restriction intensity and each pandemic and mobility variable. We conduct this analysis after dividing the time series into the identified waves of cases, with the exception of the most recent wave for which the aggregated restriction intensity remained at zero throughout its duration. The resulting correlation heatmaps are presented in Figure~\ref{fig5}, where lines indicate the pandemic and mobility variables and columns represent each wave. Positive correlations are indicated by blue shades and negative correlations by red shades, while nonsignificant correlations are shown in gray. 

With the exception of the reproduction number [$\mathcal{R}(t)$], all other variables exhibited a significant correlation with the aggregated restriction intensity during the first wave. Mobility indicators related to urban settings were strongly and negatively correlated with the restriction intensity, whereas the indicator of home permanence demonstrated a strong positive correlation. Pandemic indicators, in turn, were negatively correlated to the aggregated restriction intensity during this wave. These findings are consistent with our qualitative description of the first wave and show that the stringent measures implemented earlier led to an overall reduction in mobility throughout the city and an increase in the amount of time spent at home. However, in retrospect, these measures can be deemed overly severe, and their subsequent lifting resulted in a negative correlation with the pandemic indicators. During the second wave, only home permanence positively correlated with the restriction intensity, and all other mobility indicators were nonsignificant associated with this quantity. Among the pandemic variables, only the number of deaths positively correlated with the restriction intensity, starting a pattern that remained until the fifth wave. 

During the most severe phase of the pandemic (third and fourth wave), we observe even stronger negative correlations with mobility indicators pertaining to urban settings as well as the highest positive correlations with home permanence. These results show once again that restrictions imposed by the city administration were effective in decreasing the city mobility. Among the pandemic variables, only the occupancy of infirmary beds was not correlated with the aggregated restriction intensity, while all other variables were positively correlated. Finally, during the fifth wave, all mobility indicators were nonsignificantly correlated to the restriction intensity and only the number of cases and deaths correlated positively with this quantity.

When combined, the findings depicted in Figure~\ref{fig5} indicate that the measures implemented by the city administration had a remarkable impact on both mobility and pandemic indicators. Notably, the effects were most pronounced during the onset of the pandemic's uncertainty (first wave) and the most severe phase of the pandemic (third and fourth waves). However, positive correlations between restriction intensity and pandemic indicators were consistently observed post the first wave, suggesting that the city's measures were more reactive than proactive in response to the pandemic situation. This hypothesis is further supported by the lack of correlation between the aggregated restriction intensity and the reproduction number, which was a valuable indicator for anticipating changes in case numbers during all waves. Furthermore, we find statistically significant correlations between the reproduction number and backward-shifted aggregated restriction intensity during the third and fifth waves. Specifically, with a three-week lag between these two time series, the Spearman correlation was $0.88$ and $0.79$ ($p$-values~$< 0.001$) during the third and fifth waves, respectively. This is a critical issue because research with theoretical epidemic models has indicated that poor timing of non-pharmaceutical interventions may significantly increase transmission~\cite{lai2020effect}.

\subsection*{Comparison with cities of similar size and with the national scenario}

A cross-national hierarchical study has shown that different types of non-pharmaceutical interventions have varying impacts on controlling disease transmission.~\cite{brauner2021inferring}. Thus, and despite the lack of detailed information about other cities' measures to contain the pandemic, we compare the impact of COVID-19 in Maring\'a with five selected cities of similar size (namely: Campina Grande - PA, Piracicaba - SP, Rio Branco - AC, Santos - SP, and São Jos\'e do Rio Preto - SP), as well as with the national scenario, using the available variables. We chose cities of similar size because there are crucial differences in disease spread dynamics between small and large municipalities, as discussed in the context of Figure~\ref{fig1}B.  To compare the number of cases, we restrict our analysis to the period from the date of the first case in each city and in Brazil to 22 August 2021, which is the maximum time span for which this information was available for all cities and the country. Similarly, for deaths, we consider information up to 12 June 2022 due to the more comprehensive data obtained from the Civil Registry (see Methods).

At the level of the Brazilian states, we find that different non-pharmaceutical interventions were proposed during distinct moments (Figures~S4-S8). However, the variability related to dynamic processes of imposing and relaxing control measures is significantly higher at the city level (Figure~\ref{fig4}A). This considerable variability once again reflects the degree of autonomy that cities had during the COVID-19 crisis in Brazil~\cite{rodrigues2021brazil, biehl2021supreme}. Additionally, we observe high variability in the mobility patterns among the five similarly sized cities (Panels D in Figures~S9-S13), suggesting the timing of control measures and the measures themselves were indeed heterogeneous among these locations. 

Between January 2020 and August 2021, Brazil experienced two epidemic waves, while each of the cities in our study experienced three to four waves (Panel A in Figures~S1 and S9-S13). São Jos\'e do Rio Preto had the highest average number of weekly cases per 100,000 inhabitants ($\approx$ 229 weekly cases per 100,000 people), followed by Maring\'a with $\approx$ 192 weekly cases per 100,000 inhabitants (Table~S2). Maring\'a had 54\% more cases per 100,000 than the national average of $\approx$ 125 weekly cases per 100,000 inhabitants. In contrast, Rio Branco had the lowest number of $\approx 107$ weekly cases per 100,000 people. We further calculate the maximum number of weekly cases per 100,000 inhabitants to characterize the peak transmission in these cities and in the country (Table~S2). Again, São Jos\'e do Rio Preto had the greatest peak ($\approx$ 672 weekly cases per 100,000 people) in the epidemiological week ending on 20 June 2021, while Rio Branco displayed the lowest peak ($\approx$ 384 weekly cases per 100,000 people) in the epidemiological week ending on 7 March 2021. Maring\'a stands in between these two cities, with a maximum of 500 weekly cases per 100,000 inhabitants in the epidemiological week ending on 13 December 2020. Interestingly, all cities showed remarkably higher peaks in cases than the national number of 253 cases per 100,000 inhabitants in the epidemiological week ending on 28 March 2021. 

Focusing now on the instantaneous reproduction number $\mathcal{R}(t)$, we first estimate the maximum value of the 7-day moving average of this quantity ($\mathcal{R}_{\text{max}}$). São Jos\'e do Rio Preto exhibited the highest value among the cities in our study, with $\mathcal{R}_{\text{max}} = 2.49$ in May 2020 (Table~S2). In general, the dates of these maxima are concentrated in the initial phases of the COVID-19 spread during early 2020 (Panel B in Figures~S1 and S9-S13), as our comparative data does not span the surge of cases caused by the Omicron variant. With the exception of Santos, which displayed the smallest value of $\mathcal{R}_{\text{max}} = 1.65$ in April 2020, all other cities exhibited $\mathcal{R}_{\text{max}}>2$. Unlike the cases, the reproduction number was considerably higher for the country as whole, with $\mathcal{R}_{\text{max}} = 4.27$ occurring in March 2020. 

In addition to examining the peak behavior of the reproduction number $\mathcal{R}(t)$, it is also of interest to quantify the proportion of time during which this metric remained above the epidemic threshold ($\mathcal{R}(t) > 1$), indicating transmission acceleration. Although Maringá did not experience the highest peak in cases or $\mathcal{R}(t)$, it had the highest proportion of days with $\mathcal{R}(t)>1$ between January 2020 and August 2021. Specifically, the reproduction number remained above the epidemic threshold for 59\% of the time in Maringá, which is the same as the fraction observed across the country as a whole (see Table~S2). In contrast, Rio Branco displayed the smallest fraction, with $\mathcal{R}(t)>1$ for only 44\% of the time. Not only was the proportion of time above the epidemic threshold higher in Maring\'a, but also the average number of consecutive days with $\mathcal{R}(t)>1$ was the highest ($\approx$ 17 days). Rio Branco, once again, had the smallest average number of consecutive days above the epidemic threshold ($\approx11$ days). In turn, the national fraction was considerably higher, with approximately 19 consecutive days with $\mathcal{R}(t)>1$, possibly reflecting the effect of aggregating cases from cities in different phases of spread.

In accordance with the trends in cases, São Jos\'e do Rio Preto exhibited one of the highest number of weekly deaths ($\approx$ 3.6 weekly deaths per 100,000 people) as well as the highest maximum number of weekly deaths ($\approx$ 30 weekly deaths per 100,000 people in the epidemiological week ending on 4 April 2021, Table~S3). We also refer to Panel C in Figures~S1 and S9-S13 for visualizations of the mortality time series for all cities and the country as a whole. Notably, these figures are markedly higher than the corresponding national figures of $\approx$ 1.6 weekly deaths per 100,000 people and $\approx$~11 weekly deaths per 100,000 people at the peak week mortality in the epidemiological week ending on 28 March 2021. In contrast, Rio Branco recorded the lowest mortality figures in our comparison, with numbers similar to the national figure. Maring\'a presented $\approx$~3.7 weekly deaths per 100,000 people (the highest number of weekly deaths) and $\approx$ 23 weekly deaths per 100,000 people at peak week mortality in the epidemiological week ending on 28 March 2021. 

We further perform a demographic comparison of mortality between Maring\'a and Brazil. Specifically, we use the age and gender structures from 2010, adjusting their values based on the ratio between the population estimate of 2021 and the population in 2010, allowing us to estimate the demographic structure of each region in 2021. Although the fraction of deaths in each age demographic group was comparable, Maring\'a had a higher number of deaths per 100,000 inhabitants for all age groups above 30 years (Table~S4). This difference was most prominent in the 30-39 age group, with 43\% more deaths per 100,000 than the national rate. The situation differed when considering the gender structure. Among males, Maringá showed approximately 70\% more deaths per 100,000 people than Brazil in the 30-39 and 40-49 age groups, and mortality rates between 20\% and 35\% for older age groups. In contrast, for females, the mortality rate differences were not higher than 30\% when compared to the national figures, except for the 0-19 age group.

Finally, we investigate differences in vaccine coverage between Maring\'a and Brazil using data from the Oswaldo Cruz Foundation (Fiocruz). Maring\'a consistently exhibited a higher percentage of vaccine coverage in comparison to the national rate (Figure~S14). The maximum discrepancy of 14.12\% was observed in July 2021, while the minimum difference of 0.49\% occurred in February 2021 during the early stages of the vaccination plan.

\section*{Discussion and Conclusion}\label{sec:conclusions}

We have presented a comprehensive investigation of the impact of the COVID-19 pandemic in Maring\'a, a medium-sized city in Brazil's South Region. In contrast to previous research that has primarily described the pandemic dynamics at regional and national levels, we have focused on understanding the disease dynamics and its consequences at the city level. To overcome difficulties in obtaining local data beyond traditional epidemiological variables, we have actively monitored the pandemic in the city for 817 days between 18 March 2020 and 12 June 2022. This initiative produced a unique dataset that includes not only more traditional epidemic variables but also daily hospital bed occupancy, daily progress in vaccine coverage, and all city administration decrees that imposed and relaxed non-pharmaceutical interventions. Additionally, we have meticulously analyzed all city decrees and classified their impacts on daily life into different categories, creating a detailed timeline of all non-pharmaceutical interventions implemented in response to the pandemic. This information allowed us to create a daily indicator of the overall intensity of the restrictions imposed on citizens, which in turn proved helpful in quantifying the effect of city administration decisions to mitigate the impact of COVID-19.

We have found that the pandemic unfolded in the city through a sequence of waves characterized by significantly heterogeneous behaviors. At the outset of the pandemic, stringent non-pharmaceutical interventions were implemented due to the uncertainty surrounding the new disease, and this greatly reduced mobility in the city. Still, we have observed an exponential increase in the number of cases during the first two weeks after the first reported case. Additionally, the basic reproduction number during this early phase was quite close to the lower bound of the estimates obtained from the initial outbreak in Wuhan. These initial measures were excessively severe and were subsequently lifted, resulting in negative correlations between the aggregated restriction intensity and nearly all pandemic and mobility variables during the first wave. Our findings showed that restriction intensity once again correlated with almost all variables during the city's deadliest moment of the pandemic. However, this time, we have observed a positive association with pandemic variables and a negative association with mobility indicators (except for home permanence, which was positively correlated). These results suggest that the city's measures were more reactive than proactive to the pandemic situation, which was supported by the absence of correlation between the restriction intensity and the reproduction number. These correlations became statistically significant only after a time lag of three weeks, further corroborating that the city's measures were not anticipating changes in case numbers.

Our results revealed that the demographic profile of COVID deaths differed considerably across the waves of cases, but men were disproportionately affected and amounted to almost 60\% of all reported deaths. Individuals aged 60 years or older accounted for approximately 80\% of all deaths during the initial and subsequent waves. However, during the deadliest phase of the pandemic, the proportion of older adults who succumbed to COVID-19 decreased significantly to around 70\% during the third wave and to approximately 58\% during the fourth wave. As the immunization program progressed, despite the surge in cases linked to the Omicron variant, the number of deaths was much lower during the last two waves covered by our data set, accounting for only about 8\% of all deaths. Nonetheless, the profile of COVID-19 deaths reversed during the fifth and sixth waves, with older adults representing more than 90\% of all deaths during these waves.

We have also identified distinct periods during which excess mortality in the city cannot be solely attributed to deaths caused by COVID-19. In March 2020, during the first wave, there was an excess mortality of 139 individuals, while only two people died from COVID-19 that month. Although it is difficult to determine the definitive cause of this behavior, we have suggested that possible explanations may include underreporting, earlier circulation of the virus in the city (as observed at the country level~\cite{candido2020evolution, stringari2021covert}), and avoidance of medical care. We have verified that the peak in excess mortality coincided with the most severe phase of the pandemic in the city. However, even when projecting the peak of mortality caused by COVID-19 for an entire month, one cannot fully account for the excess mortality of 486 individuals in March 2021. Underreporting of deaths may have once again contributed to this figure, but the overloaded healthcare system, which remained at full capacity for almost six weeks during this period, likely played a critical role in increasing mortality due to other causes. Furthermore, we have identified an excess mortality of approximately 100 individuals per month from the fifth wave on January 2022. Although the fifth and sixth waves registered 153 deaths caused by COVID-19, there was an excess mortality of 630 individuals during these waves. As in the first wave, explaining the excess mortality during these two last waves is also challenging. We have suggested that the explosive number of cases during this period may have increased underreported deaths. However, more research is necessary to fully understand this effect, and deaths indirectly related to COVID-19 and its aftereffects may play a role in this phenomenon.

Finally, despite the lack of similarly detailed information for other cities, we have conducted a comparative analysis between the impact of COVID-19 in Maring\'a and in five other selected cities of similar size, as well as the national scenario. These results revealed a high variability in the mobility patterns and the evolution of epidemiological variables, indicating that both the timing and intensity of control measures were heterogeneous among these locations. All these heterogeneities can be partially attributed to the vast geographic extension of Brazil and the nonconcurrent patterns of spread~\cite{castro2021reduction, candido2020evolution}. However, we have further emphasized that Brazilian cities had an unusual and noteworthy role in managing the COVID crisis, following the Supreme Court's decision against the centralization of decision-making by the federal government. In addition to socioeconomic disparities among cities, political issues likely played a role in creating this heterogeneous and complex scenario, as cooperation and coordination between different government levels were hindered during the COVID crisis~\cite{de2022quantifying}. 

To conclude, we hope our work emphasizes the importance of individually investigating the pandemic at the level of cities. By doing so, we can better understand the dynamics of infectious diseases and develop more targeted and effective control measures. We further believe our findings have implications for policymakers and public health officials in Brazil and other countries with large urban centers. We hope that our results trigger similar initiatives for actively monitoring the spread of other infectious diseases and corresponding city measures to mitigate their impacts, as well as possible comparative studies about the COVID-19 pandemic at a similar level of detail. 

\section*{Methods}

\subsection*{Data}

The datasets used in our study were obtained from various sources. The primary source of information originated from our initiative to actively monitor the COVID-19 pandemic in Maring\'a (\href{https://complex.pfi.uem.br/covid}{\textit{Observat\'orio COVID-19 Maring\'a}}). Specifically, we have collected and processed each of the daily reports published by the city administration between 18 March 2020 and 12 June 2022~\cite{notifica_saude}. From these daily reports, we have extracted the number of COVID cases and deaths, the occupancy of hospital beds (both infirmary and ICU) exclusive to COVID-19 patients and for general use, as well as the evolution of vaccine coverage. Additionally, using information from these daily reports, we have gathered all city decrees~\cite{legislacao-municipal} that imposed and relaxed non-pharmaceutical interventions implemented to mitigate the spread of COVID-19. These decrees were scrutinized and their restrictions were classified into the categories shown in Figure~\ref{fig4}A.

We have also used the Johns Hopkins Coronavirus Resource Center to obtain the number of COVID-19 cases at the national level~\cite{ourworldindata_covid} and the \textit{brasil.io} API~\cite{brasil_io} to gather the same information for all the 5,570 Brazilian cities. Missing data were filled by linearly interpolating the datapoints. The open-source \textit{brasil.io} API gathered information from daily reports published by the Health Offices of each of the 26 states and the federal district. To compare the vaccination coverage between Maring\'a and Brazil, we have obtained national-level data from the online dashboard \textit{MonitoraCovid-19}, provided by the Oswaldo Cruz Foundation (Fiocruz)~\cite{fiocruz-vaccination}. To calculate the rates of cases and deaths per 100,000 inhabitants, we have used city population estimates for 2021 released by the Brazilian Institute of Geography and Statistics (IBGE)~\cite{ibge_2021}. We have further used data from the open source Oxford COVID-19 Government Response Tracker (OxCGRT) project~\cite{oxford_response_tracker} to obtain non-pharmaceutical intervention data at state level. 

The Civil Registry of Natural Persons was our primary source for COVID-19 mortality data~\cite{civil-registry}. This unified national data source relies on documents registered by local notary offices at the national Civil Registry Information Center (CRC). The civil registry dataset includes mortality numbers for all natural causes of death, as well as demographic data (age and gender) for all Brazilian cities. Our data set covers the period from 1 January 2019 to 12 June 2022. We have calculated the monthly number of excess deaths for 2020, 2021, and 2022 by subtracting the monthly numbers from the monthly deaths in 2019. In addition, we have used Google's community mobility reports~\cite{google-mobility} to estimate the changes in mobility between 25 February 2020 and 12 June 2022. This metric reflects the percentage change in the number of visits and length of stay compared to the median baseline value between 3 January 2020 and 6 February 2020, before the emergence of COVID-19. Google's community mobility reports are derived from anonymous aggregated data used in the ``popular times'' feature of Google Maps. The six data streams that gauge mobility include grocery and pharmacy, parks, transit stations, retail and recreation, residential, and workplaces. The mobility data enables us to infer the impact of social distancing measures on reducing the spread of COVID-19.

The data necessary to replicate our results is public available at the GitHub repository \url{https://github.com/ansesu/covid19_maringa}.

\subsection*{Estimating the initial exponential growth rate of cases}

Epidemics often exhibit an increasing exponential trend in their initial phase~\cite{picoli2011spreading}. For new diseases or variants of existing ones, the entire population is susceptible and infected individuals can spread the disease to the whole population. Mathematically, an exponential increase in the number of infected individuals can be expressed as
\begin{equation}\label{eq:exp}
    I(t) = I_0 \exp(rt)
\end{equation}
where $I(t)$ correspond to the number of cases of COVID-19 at time $t$ (measured in days since the first reported case), $I_0$ is the initial number of cases, and $r$ represents the exponential rate of growth. We have adjusted the linearized version of Eq.~\ref{eq:exp}, that is,
\begin{equation}\label{eq:exp_lin}
    \log I(t) = \log I_0 + rt\,,
\end{equation}
to the initial evolution of the number of COVID-19 cases in all Brazilian cities using the Python package \textit{statsmodels}~\cite{seabold2010statsmodels}.

\subsection*{Estimating the basic and instantaneous reproduction numbers}

In order to provide a more precise description of the evolution of an epidemic, it is a common practice to estimate the basic reproduction number $\mathcal{R}_0$, which measures the strength of the spread~\cite{dushoff2021speed}. This metric represents the average number of infections caused by a typical infected individual during the initial stage of an epidemic, when the entire population is susceptible.

We have calculated the basic reproduction number using the susceptible-infected-recovered (SIR) model~\cite{kermack1927contribution}, defined by the following set of differential equations:
\begin{equation}
    \begin{split}
    \dfrac{dS(t)}{dt} &= -\beta \dfrac{S(t)}{N}I(t)\,,\\ 
    \dfrac{dI(t)}{dt} &= \beta \dfrac{S(t)}{N}I(t) - \gamma I(t)\,,\\
    \dfrac{dR(t)}{dt} &= \gamma I(t)\,,
    \label{eq:sir}
    \end{split}
\end{equation}
where $S(t)$ is the number of susceptible individuals at time $t$, $R(t)$ is the number of recovered individuals at time $t$, $N$ is the total population, $\beta$ is the contact rate, and $1/\gamma$ is the infectious period. In the initial stage of a pandemic, the number of initially susceptible individuals can be approximated by the city population ($S \approx N$). Under this assumption, the number of infected individuals in Eq.~\ref{eq:sir} can be written as:
\begin{equation}\label{eq:r0_partial}
\dfrac{dI(t)}{dt} = \gamma (\mathcal{R}_0 - 1) I\,,
\end{equation}
where $\mathcal{R}_0 = \beta/\gamma$. By comparing the solution of the previous equation with Eq.~\ref{eq:exp}, the basic reproduction number can be written as:
\begin{equation}\label{eq:r0}
    \mathcal{R}_0 = 1+r/\gamma\,.
\end{equation}
It should be noted that the number of cases will increase exponentially if $\mathcal{R}_0>1$, whereas the number of cases will decrease exponentially if $\mathcal{R}_0<1$.

As the pandemic progresses, the assumption $S \approx N$ is no longer valid because the population of susceptible individuals becomes smaller than the total population due to immunization. In this case, we have focused on the instantaneous reproduction number $\mathcal{R}(t)$. This quantity represents the average number of infections caused by a typical infected individual at a time $t$. In practice, the number of cases increases at time $t$ if $\mathcal{R}(t)>1$ and decreases if $\mathcal{R}(t)<1$. We have estimated the instantaneous reproduction number using the Bayesian approach developed by Cori \textit{et al.}~\cite{cori2013new}, as implemented in their \textit{R} package \textit{EpiEstim}. 

Mathematically, the instantaneous reproduction number can be expressed as
\begin{equation}\label{eq:r}
    \mathcal{R}(t) = \dfrac{I(t)}{\sum_{s=1}^{\tau} I(t-s)w(s)}\ ,
\end{equation}
where $w(s)$ represents the infectivity profile in a given time $s$ and $\tau$ is the size of the time window used to calculate $\mathcal{R}(t)$. The infectivity profile is a probability distribution that indicates the infectiveness of the individual at a time $s$. In this study, we have used the serial interval distribution (time between the onset of symptoms in primary and secondary cases) as our infectivity profile (gamma distribution with parameters $\mu =4.8$ days and $\sigma = 2.3$ days~\cite{nishiura2020serial}). To account for the uncertainty in the serial interval distribution, we have sampled 500 pairs of means and standard deviations from truncated normal distributions and then estimated the 95\% confidence intervals. We have further used early estimates of the basic reproduction number in Wuhan as parameters of the prior distribution (gamma distribution with parameters $\mu = 2.6$ days and $\sigma = 2.0$ days~\cite{abbott2020transmissibility}) and set $\tau = 7$ days. Moreover, following Cori \textit{et al.}~\cite{cori2013new}, we have started estimating $\mathcal{R}(t)$ after the first twelve COVID-19 cases. 

\section*{Data availability}
The data necessary to replicate our results is public available at the GitHub repository \url{https://github.com/ansesu/covid19_maringa}.

\section*{Acknowledgements}
The authors acknowledge the support of the Coordena\c{c}\~ao de Aperfei\c{c}oamento de Pessoal de N\'ivel Superior (CAPES), the Conselho Nacional de Desenvolvimento Cient\'ifico e Tecnol\'ogico (CNPq -- Grant 303533/2021-8), and the Slovenian Research Agency (Grants J1-2457 and P1-0403). 

\section*{Author contributions statement}
A.S.S, A.A.B.P., M.P., and H.V.R. designed research, performed research, analyzed data, and wrote the paper.

\section*{Competing Interests}
The authors declare no competing interests.

\bibliography{references}

\clearpage


\section*{Supplementary material}

# Complexity of the COVID-19 pandemic in Maringá

**Andre S. Sunahara**[1], **Arthur A. B. Pessa**[1], **Matjaž Perc**[2,3,4,5,6,*], **and Haroldo V. Ribeiro**[1,†]

[1]Departamento de Física, Universidade Estadual de Maringá - Maringá, PR 87020-900, Brazil
[2]Faculty of Natural Sciences and Mathematics, University of Maribor, Koroška cesta 160, 2000 Maribor, Slovenia
[3]Department of Medical Research, China Medical University Hospital, China Medical University, Taichung, Taiwan
[4]Alma Mater Europaea, Slovenska ulica 17, 2000 Maribor, Slovenia
[5]Department of Physics, Kyung Hee University, 26 Kyungheedae-ro, Dongdaemun-gu, Seoul, Republic of Korea
[6]Complexity Science Hub Vienna, Josefstädterstraße 39, 1080 Vienna, Austria
[*]email: matjaz.perc@gmail.com
[†]email: hvr@dfi.uem.br

**Supplemental Materials**

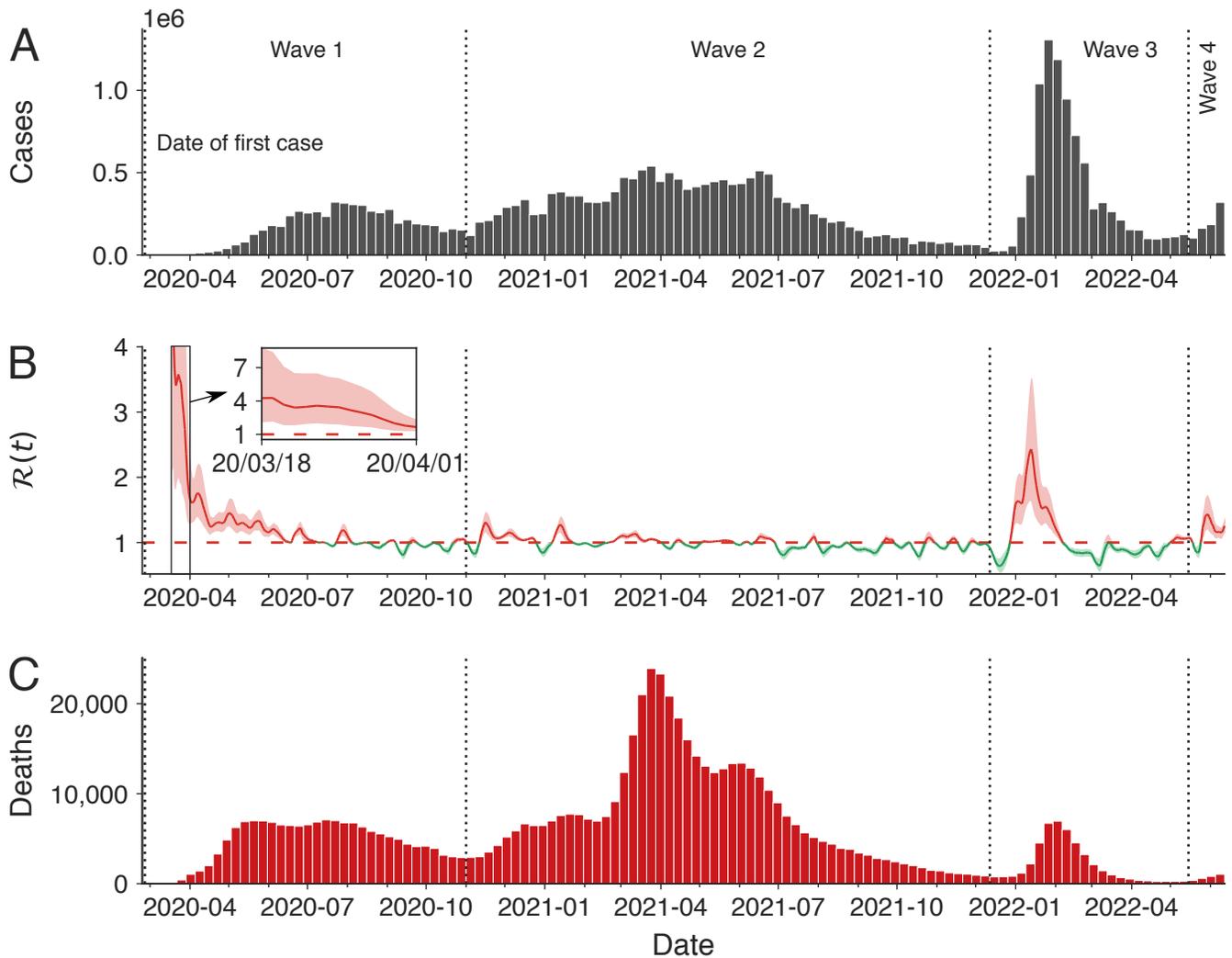

**Figure S1.** Indicators of the COVID-19 pandemic in Brazil. (A) Weekly numbers of confirmed cases of COVID-19 between 25 February 2020 and 12 June 2022. (B) Instantaneous reproduction number $\mathcal{R}(t)$ from 18 March 2020 to 12 June 2022. Shaded regions represent the 95% confidence intervals, and the dashed horizontal line indicates the epidemic threshold $\mathcal{R}(t) = 1$. (C) Weekly COVID-19 death toll between 25 February 2020 and 12 June 2022. In all panels, vertical dashed lines delineate the for identified waves of COVID-19 cases.

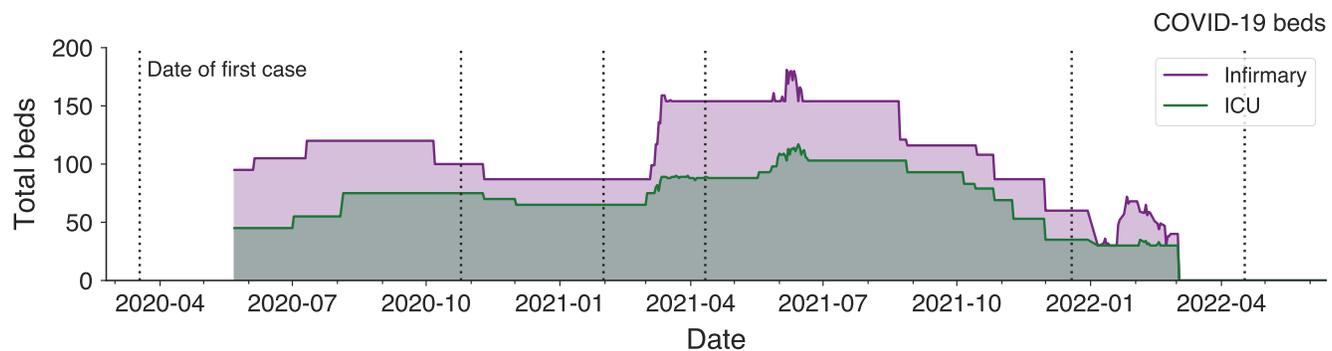

**Figure S2.** Evolution of the total number of hospital beds exclusively dedicated to COVID. Daily number of the total infirmary (purple curve) and intensive care (green curve) beds exclusively available for COVID-19 patients in Maringá from 22 May 2020 to 12 June 2022. Vertical dashed lines delineate the six identified waves of COVID-19 cases.

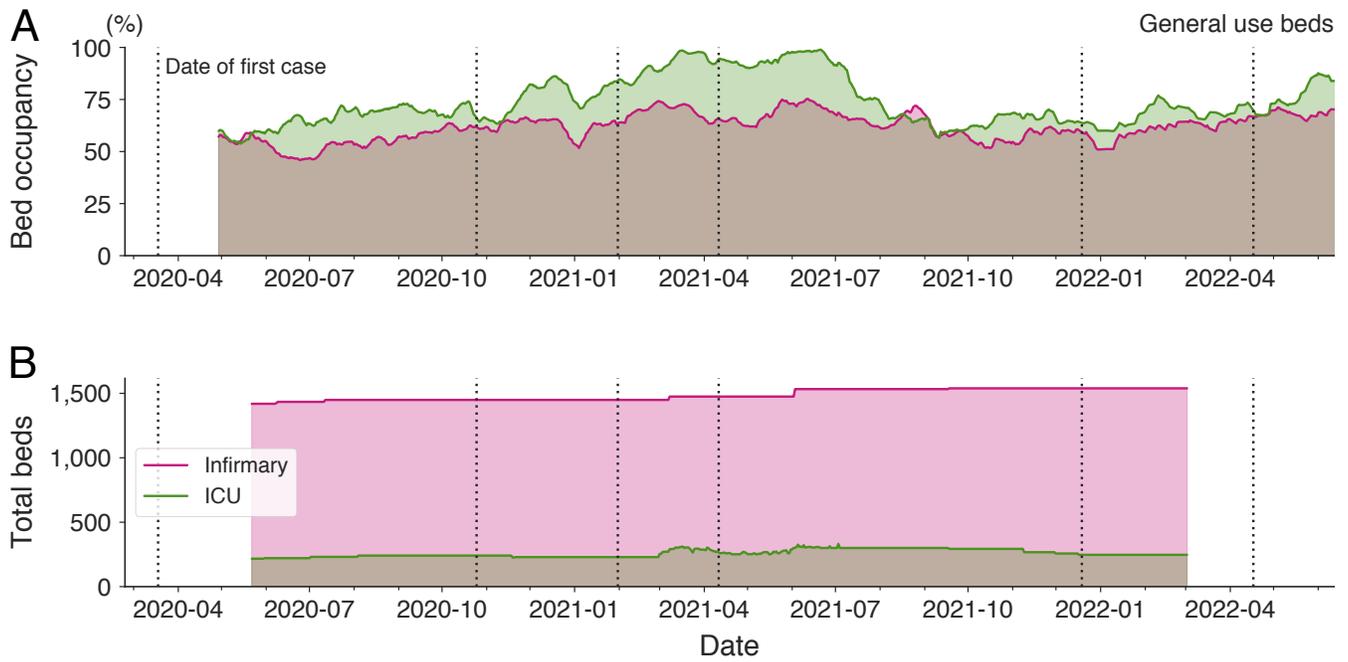

**Figure S3.** Occupancy infirmary and intensive care beds for general use. (A) Daily percentage occupancy of infirmary (purple curve) and intensive care (green curve) beds available for general use in Maringá from 29 April 2020 to 12 June 2022. The curves correspond to 7-day moving averages. (B) Daily number of the total infirmary (purple curve) and intensive care (green curve) beds for general use in Maringá from 22 May 2020 to 2 March 2022. In both panels, vertical dashed lines delineate the six identified waves of COVID-19 cases.

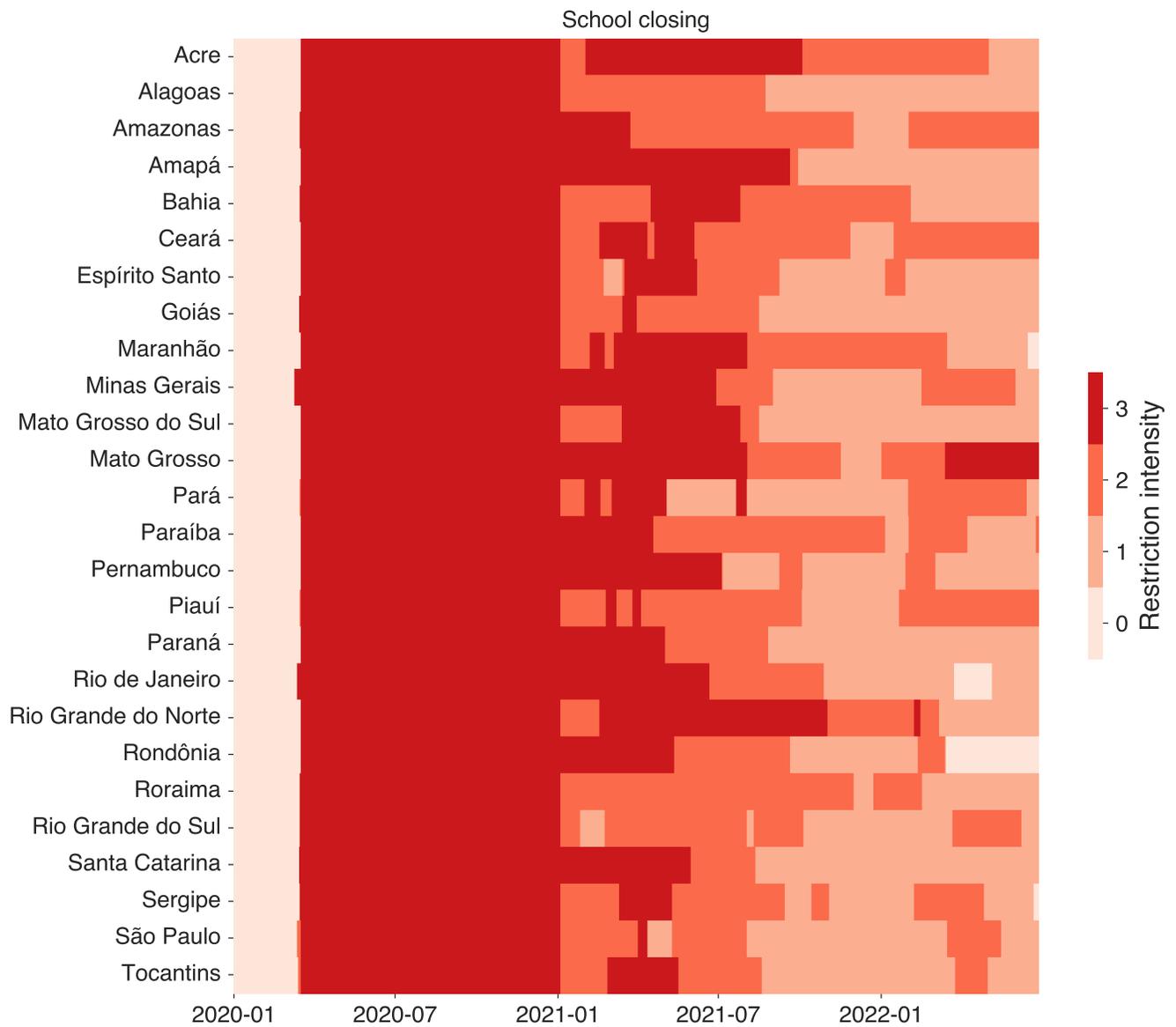

**Figure S4.** School closing non-pharmaceutical interventions among the Brazilian states. The heatmap displays the temporal evolution of the implementation of school closing interventions for all Brazilian states from January 2020 to June 2022. The color code stands for the degree of restriction: "no measures" (0), "recommend closing or all schools open with alterations resulting in significant differences compared to non-COVID-19 operations" (1), "require closing (only some levels or categories, such as just high school or just public schools)" (2), and "require closing all levels" (3).

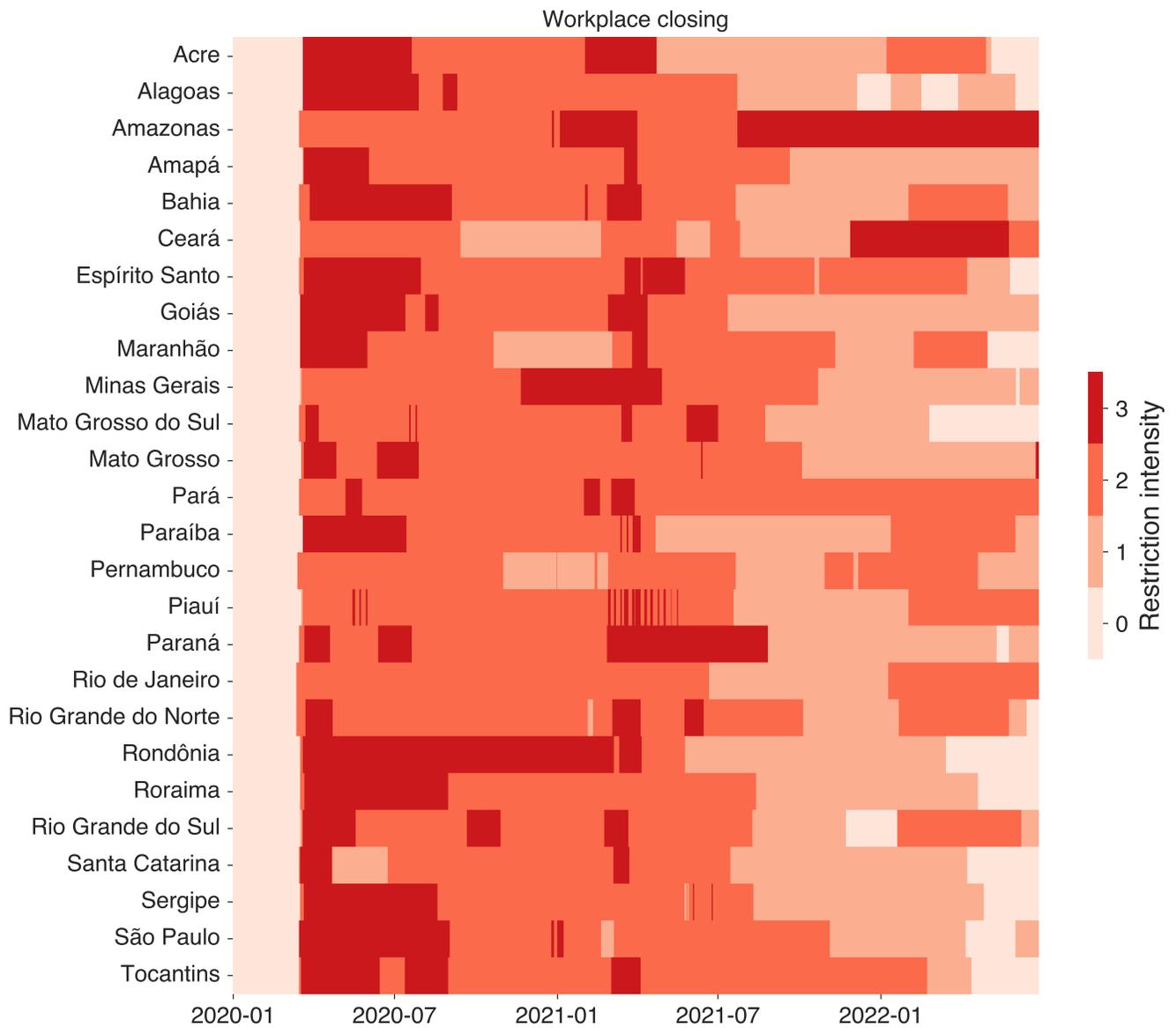

**Figure S5.** Workplace closing non-pharmaceutical interventions among the Brazilian states. The heatmap displays the temporal evolution of the implementation of workplace closing interventions for all Brazilian states from January 2020 to June 2022. The color code stands for the degree of restriction: "no measures" (0), "recommend closing (or recommend work from home) or all businesses open with alterations resulting in significant differences compared to non-COVID-19 operation" (1), "require closing (or work from home) for some sectors or categories of workers" (2), and "require closing (or work from home) for all-but-essential workplaces (e.g., grocery stores and doctors)" (3).

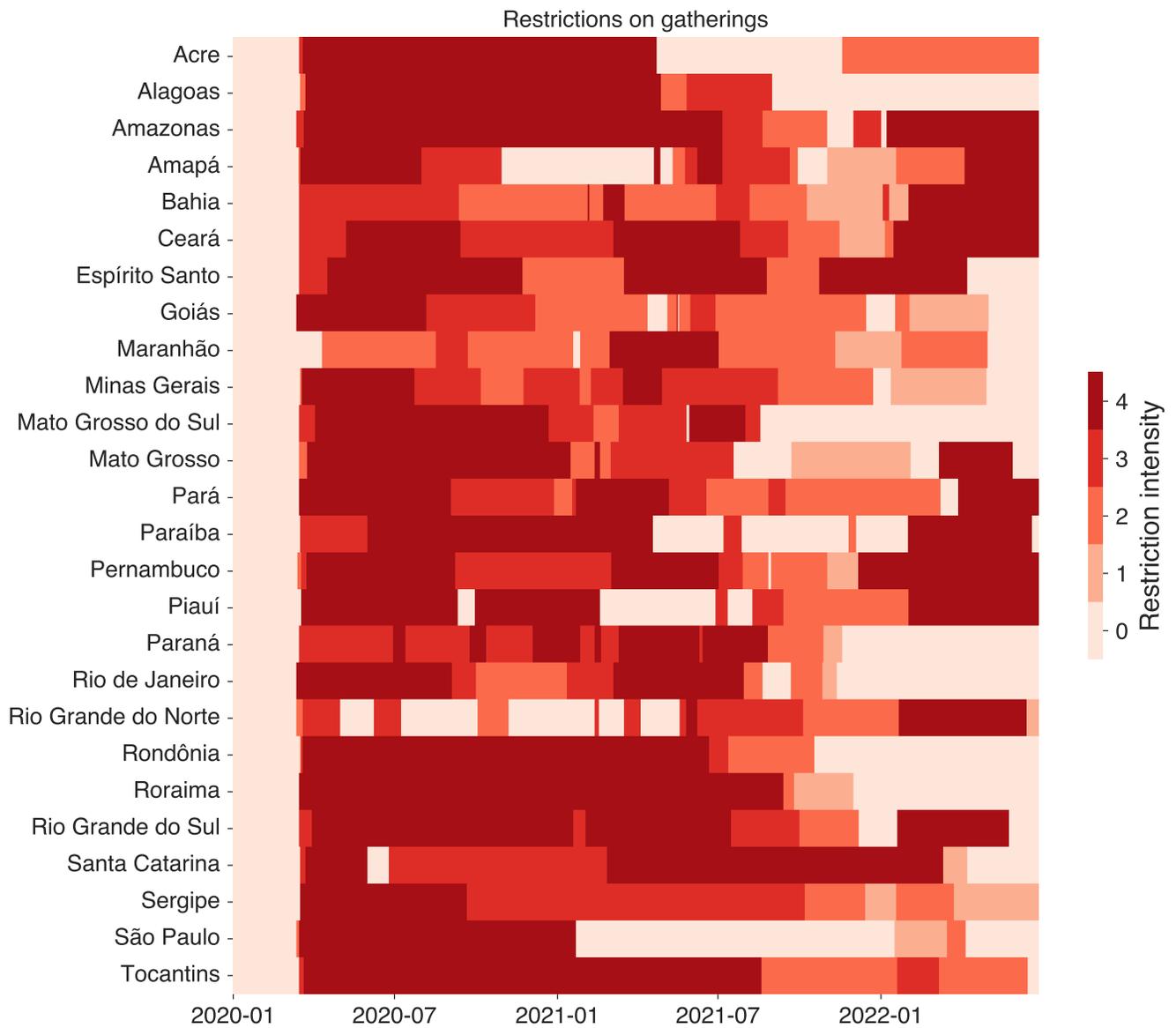

**Figure S6.** Restrictions on gatherings among the Brazilian states. The heatmap displays the temporal evolution of the implementation of restriction on gatherings interventions for all Brazilian states from January 2020 to June 2022. The color code stands for the degree of restriction: "no restrictions" (0), "restrictions on very large gatherings (the limit is above 1000 people)" (1), "restrictions on gatherings between 101-1000 people" (2), "restrictions on gatherings between 11-100 people" (3), and "restrictions on gatherings of 10 people or less" (4).

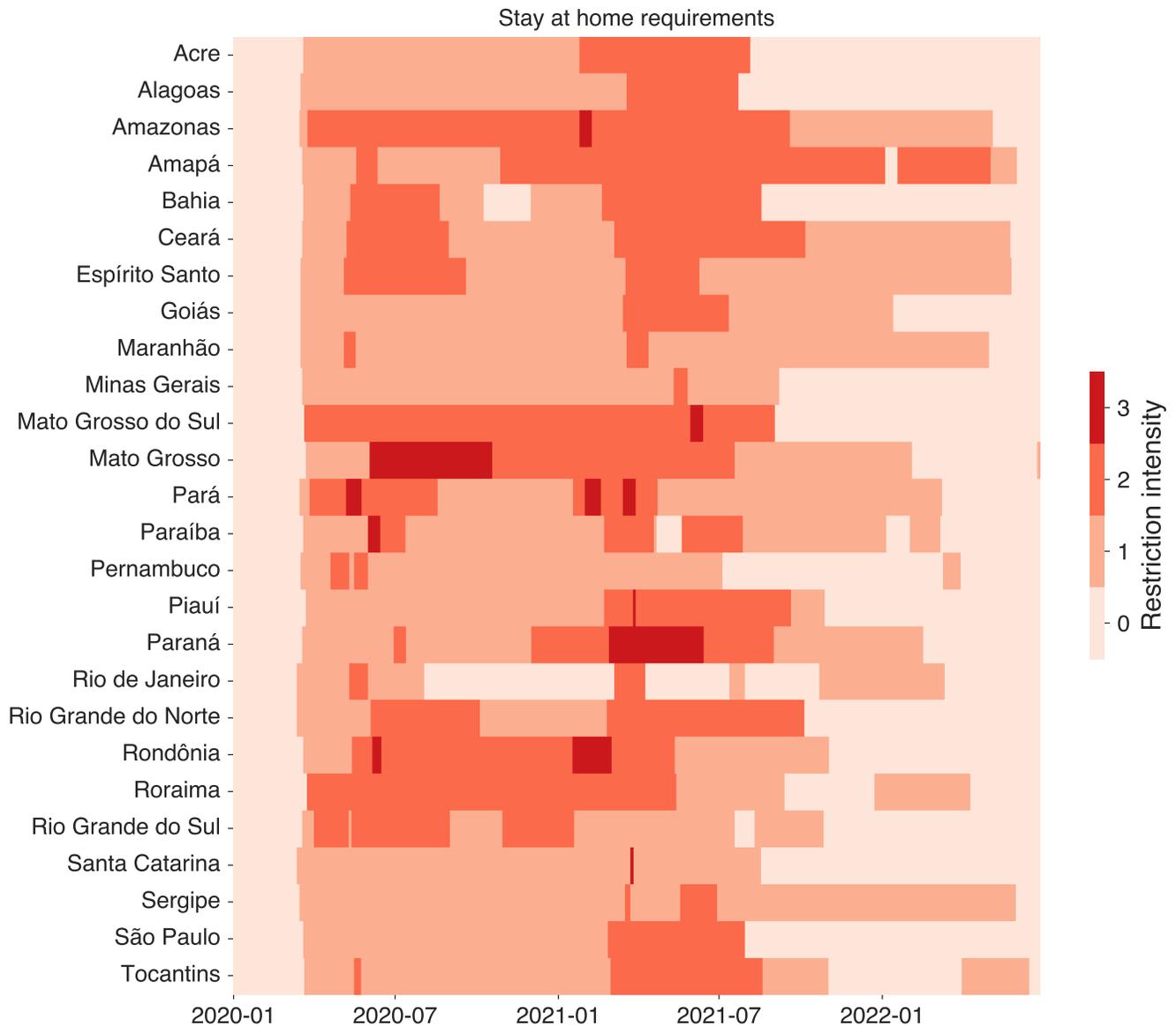

**Figure S7.** Stay at home requirements among the Brazilian states. The heatmap displays the temporal evolution of the implementation of stay at home interventions for all Brazilian states from January 2020 to June 2022. The color code stands for the degree of restriction: "no measures" (0), "recommend not leaving house" (1), "require not leaving house with exceptions for daily exercise, grocery shopping, and 'essential' trips" (2), and "require not leaving house with minimal exceptions (e.g., allowed to leave once a week or only one person can leave at a time)" (3).

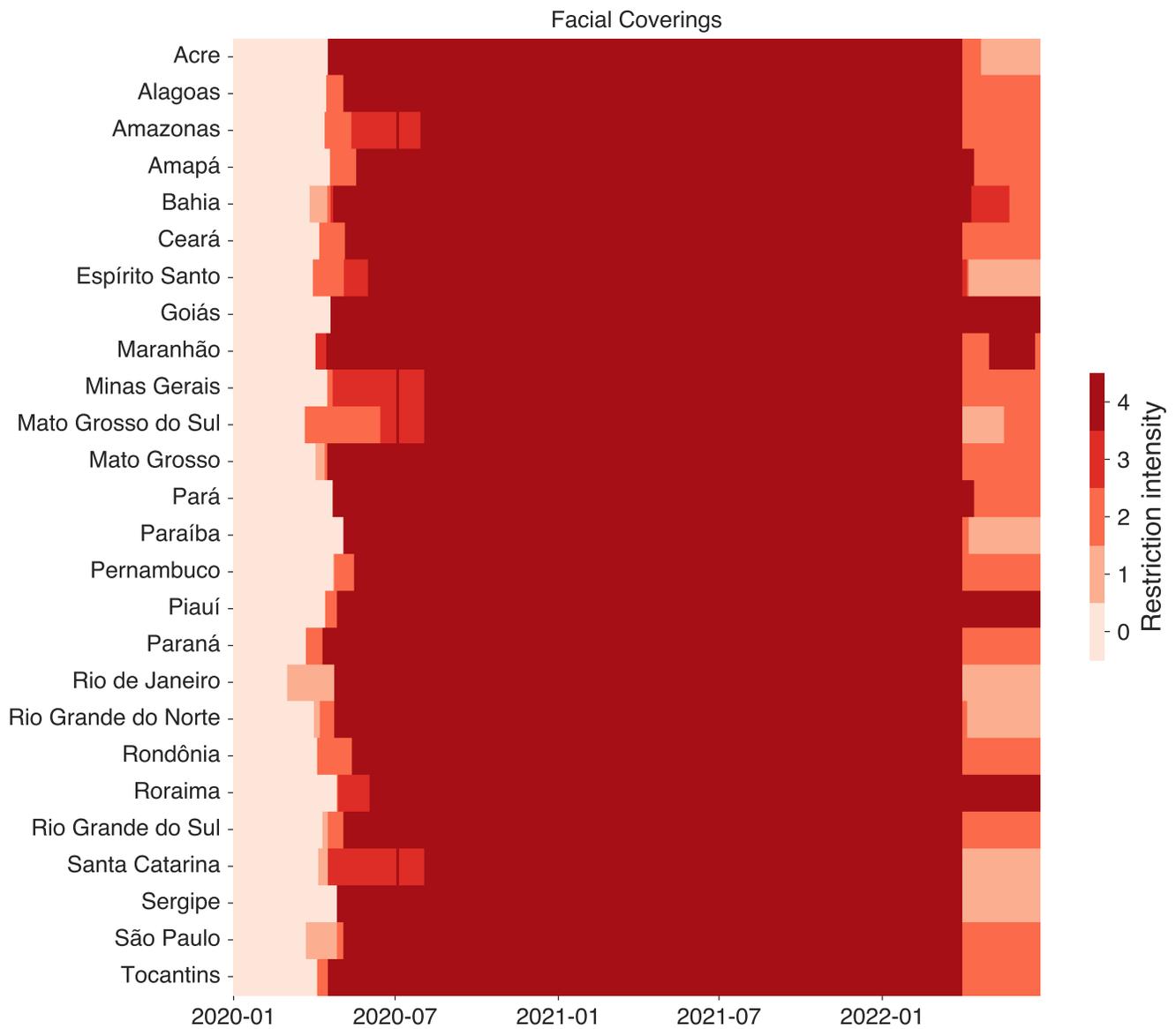

**Figure S8.** Facial covering non-pharmaceutical interventions among the Brazilian states. The heatmap displays the temporal evolution of the implementation of facial coverings interventions for all Brazilian states from January 2020 to June 2022. The color code stands for the degree of restriction: "no policy" (0), "recommended" (1), "required in some specified shared/public spaces outside the home with other people present, or some situations when social distancing not possible" (2), "required in all shared/public spaces outside the home with other people present or all situations when social distancing not possible" (3), and "required outside the home at all times regardless of location or presence of other people" (4).

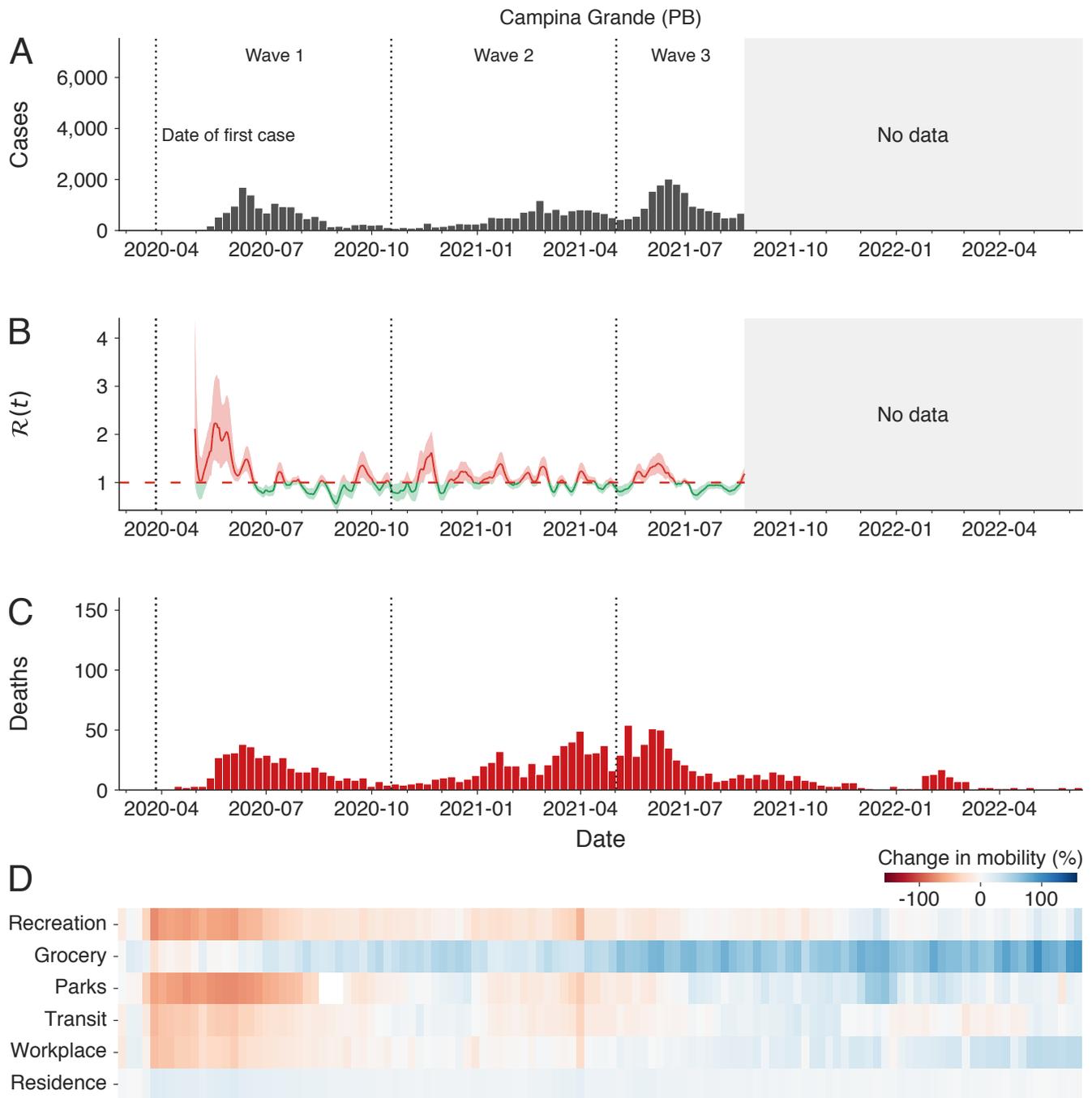

**Figure S9.** Indicators of the COVID-19 pandemic in Campina Grande (PB). (A) Weekly number of confirmed cases of COVID-19 between 22 March 2020 and 22 August 2021. (B) Instant reproduction number $\mathcal{R}(t)$ from 30 April 2020 to 22 August 2021. Shaded regions represent the 95% confidence intervals, and the dashed horizontal line indicates the epidemic threshold $\mathcal{R}(t) = 1$. (C) Weekly COVID-19 death toll between 22 March 2020 and 12 June 2022. In the previous panels, vertical dashed lines delineate the identified waves of COVID-19 cases. (D) Temporal heatmap illustrating the changes in mobility related to Google users' visiting patterns to places categorized into six groups (recreation, grocery, parks, transit, workplace, and residence) compared to baselines estimated using pre-pandemic levels. Blue shades indicate an increase in the visitation to a place category, while red shades indicate a reduction.

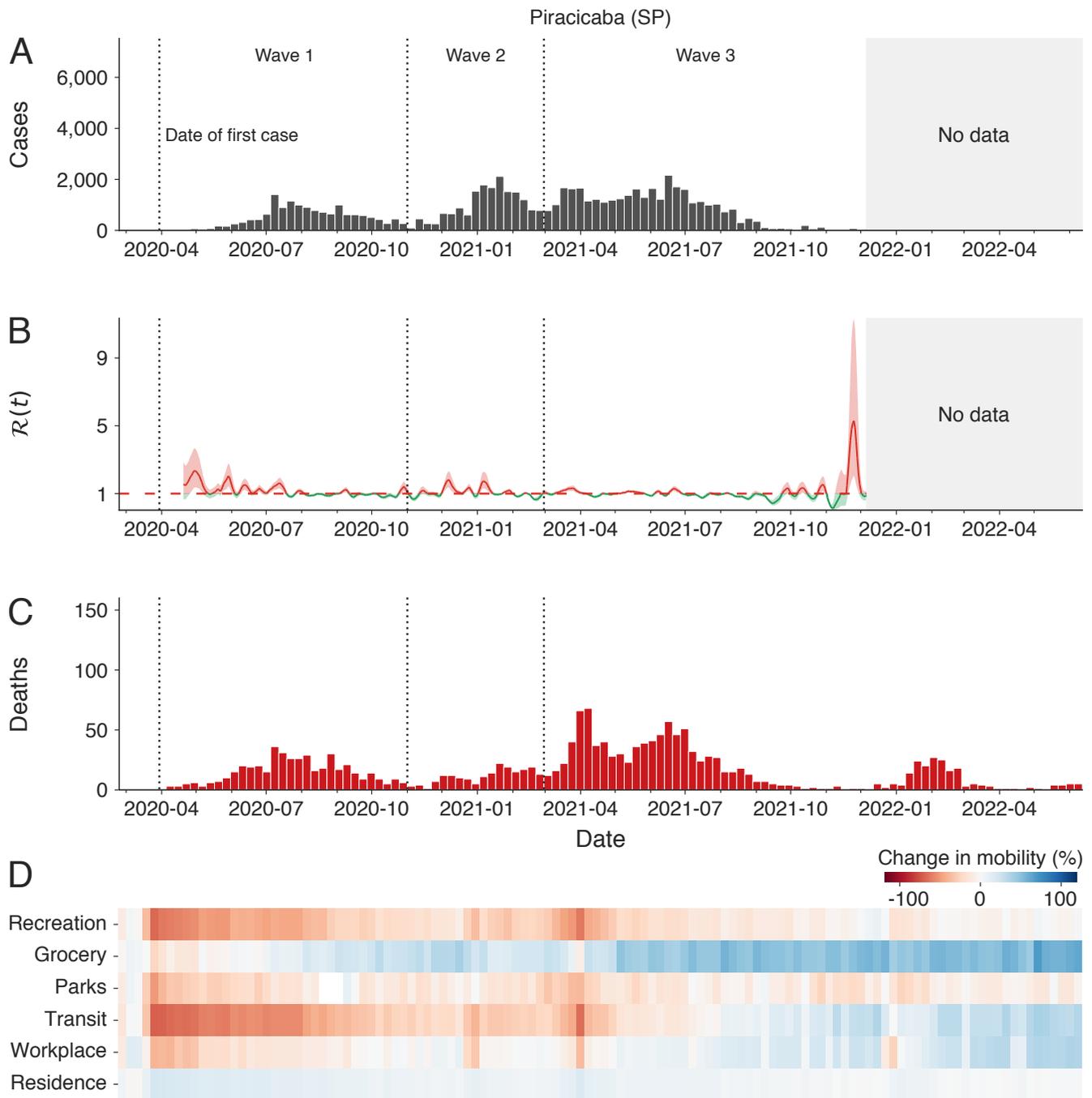

**Figure S10.** Indicators of the COVID-19 pandemic in Piracicaba (SP). (A) Weekly number of confirmed cases of COVID-19 between 29 March 2020 and 12 December 2021. (B) Instant reproduction number $\mathcal{R}(t)$ from 20 April 2020 to 06 December 2021. Shaded regions represent the 95% confidence intervals, and the dashed horizontal line indicates the epidemic threshold $\mathcal{R}(t) = 1$. (C) Weekly COVID-19 death toll between 29 March 2020 and 12 June 2022. In the previous panels, vertical dashed lines delineate the identified waves of COVID-19 cases. (D) Temporal heatmap illustrating the changes in mobility related to Google users' visiting patterns to places categorized into six groups (recreation, grocery, parks, transit, workplace, and residence) compared to baselines estimated using pre-pandemic levels. Blue shades indicate an increase in the visitation to a place category, while red shades indicate a reduction.

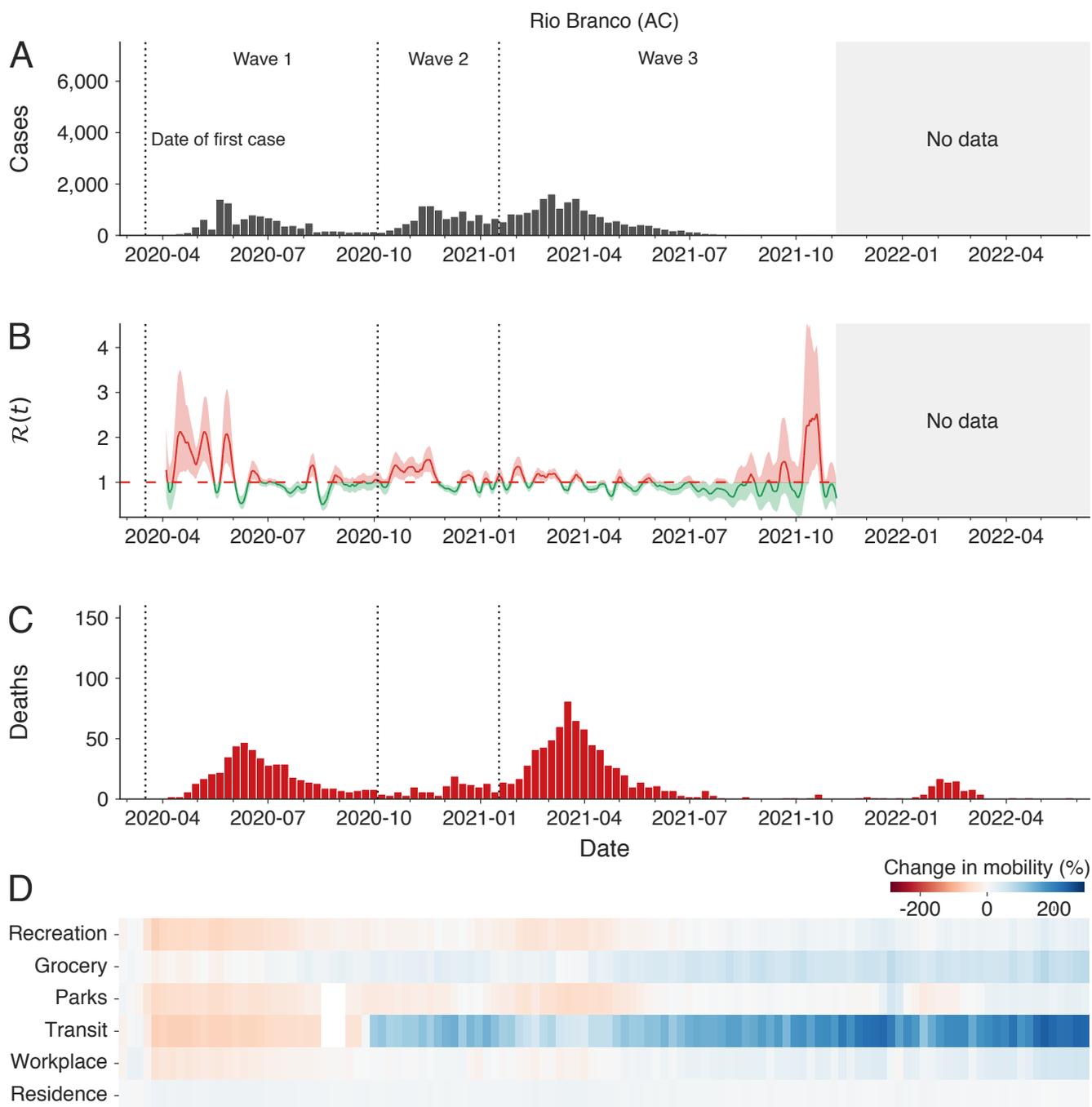

**Figure S11.** Indicators of the COVID-19 pandemic in Rio Branco (AC). (A) Weekly number of confirmed cases of COVID-19 between 15 March 2020 and 7 November 2021. (B) Instant reproduction number $\mathcal{R}(t)$ from 04 April 2020 to 05 November 2021. Shaded regions represent the 95% confidence intervals, and the dashed horizontal line indicates the epidemic threshold $\mathcal{R}(t) = 1$. (C) Weekly COVID-19 death toll between 15 March 2020 and 12 June 2022. In the previous panels, vertical dashed lines delineate the identified waves of COVID-19 cases. (D) Temporal heatmap illustrating the changes in mobility related to Google users' visiting patterns to places categorized into six groups (recreation, grocery, parks, transit, workplace, and residence) compared to baselines estimated using pre-pandemic levels. Blue shades indicate an increase in the visitation to a place category, while red shades indicate a reduction.

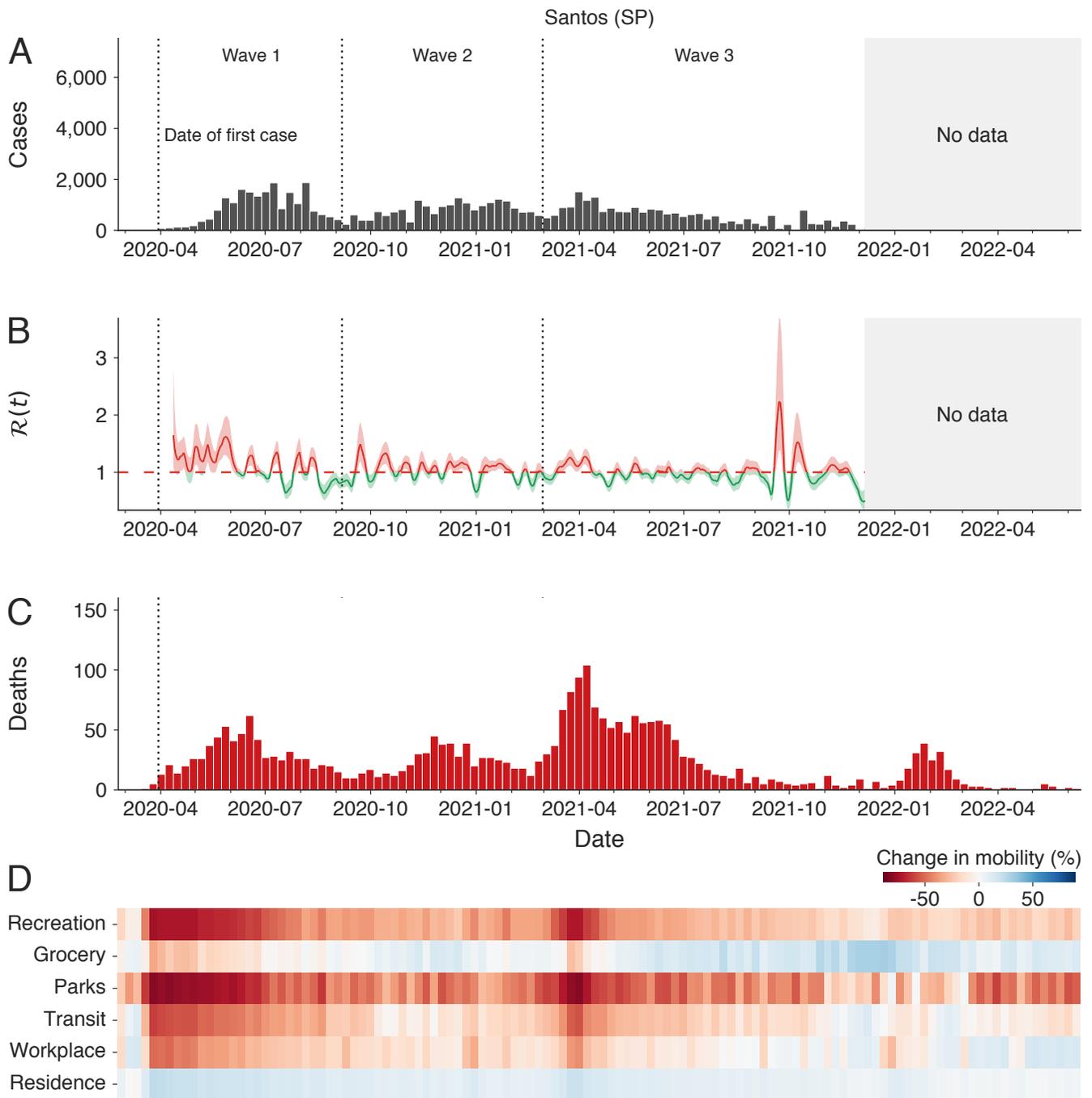

**Figure S12.** Indicators of the COVID-19 pandemic in Santos (SP). (A) Weekly number of confirmed cases of COVID-19 between 29 March 2020 and 12 December 2021. (B) Instant reproduction number $\mathscr{R}(t)$ from 12 April 2020 to 06 December 2021. Shaded regions represent the 95% confidence intervals, and the dashed horizontal line indicates the epidemic threshold $\mathscr{R}(t) = 1$. (C) Weekly COVID-19 death toll between 29 March 2020 and 12 June 2022. In the previous panels, vertical dashed lines delineate the identified waves of COVID-19 cases. (D) Temporal heatmap illustrating the changes in mobility related to Google users' visiting patterns to places categorized into six groups (recreation, grocery, parks, transit, workplace, and residence) compared to baselines estimated using pre-pandemic levels. Blue shades indicate an increase in the visitation to a place category, while red shades indicate a reduction.

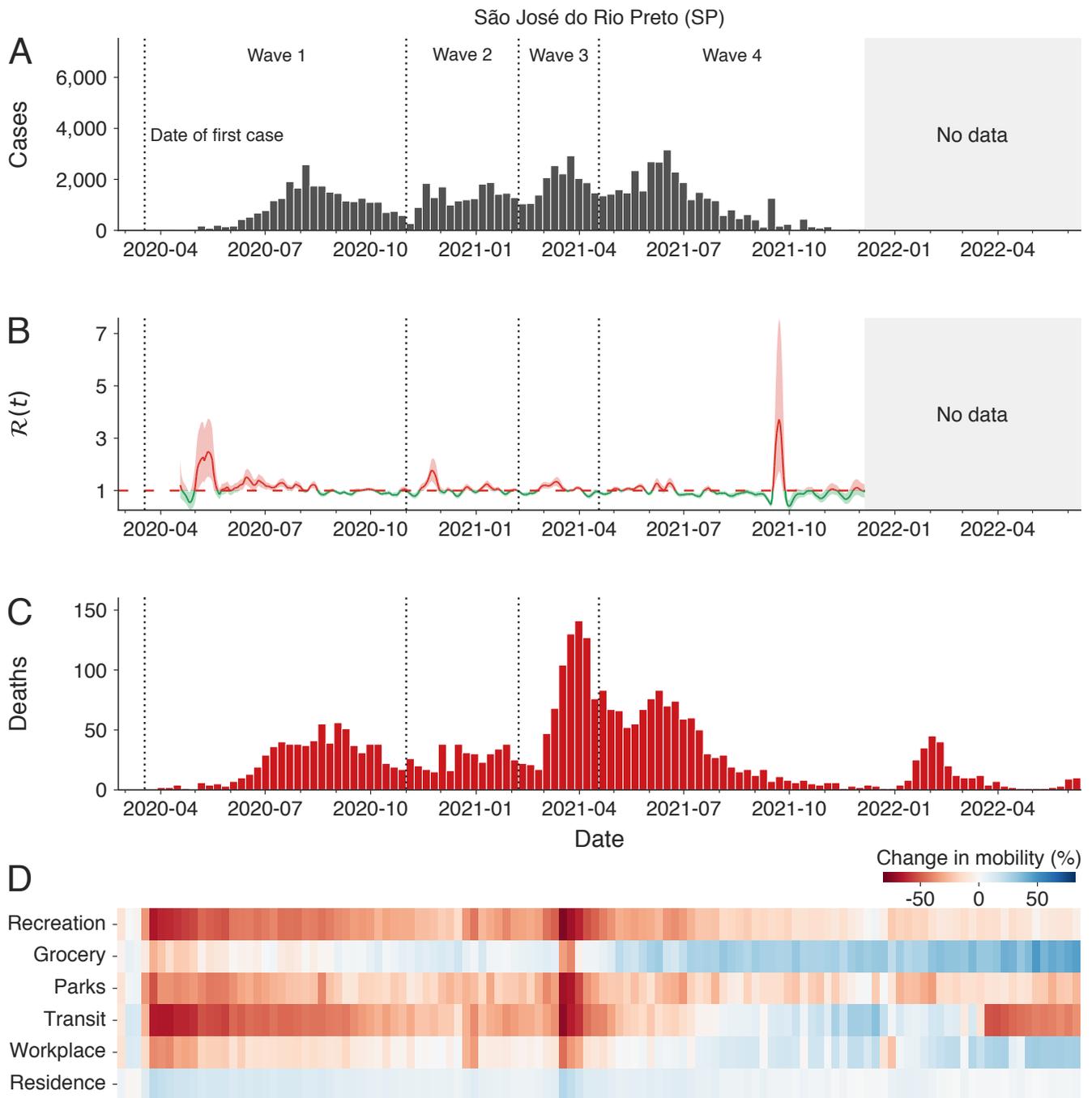

**Figure S13.** Indicators of the COVID-19 pandemic in São José do Rio Preto (SP). (A) Weekly number of confirmed cases of COVID-19 between 5 March 2020 and 12 December 2021. (B) Instant reproduction number $\mathcal{R}(t)$ from 18 April 2020 to 06 December 2021. Shaded regions represent the 95% confidence intervals, and the dashed horizontal line indicates the epidemic threshold $\mathcal{R}(t) = 1$. (C) Weekly COVID-19 death toll between 15 March 2020 and 12 June 2022. In the previous panels, vertical dashed lines delineate the identified waves of COVID-19 cases. (D) Temporal heatmap illustrating the changes in mobility related to Google users' visiting patterns to places categorized into six groups (recreation, grocery, parks, transit, workplace, and residence) compared to baselines estimated using pre-pandemic levels. Blue shades indicate an increase in the visitation to a place category, while red shades indicate a reduction.

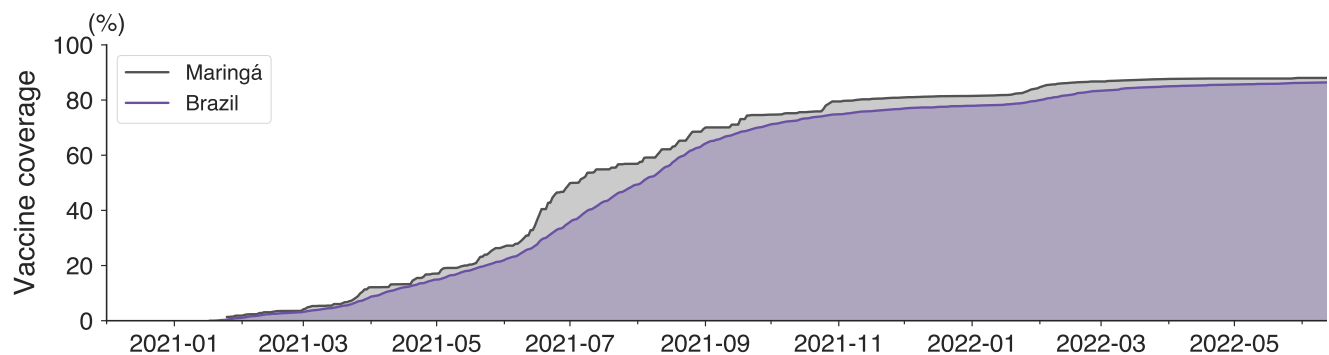

**Figure S14.** Comparison of the evolution of the vaccine coverage between Maringá and Brazil. The two curves represent the percentage of population immunized with the first dose of a COVID-19 vaccine in Maringá (gray curve) and in Brazil (purple curve) from the beginning of 2021 to 12 June 2022.

**Table S1.** Deaths and percentage of deaths caused by COVID-19 in each epidemic wave in Maringá. The table shows information about the deaths and percentage of deaths grouping the population by age and gender in each epidemic wave that took place in city. The percentage of deaths in each wave may not sum up to a hundred per cent because the total number of casualties can include deaths classified as undefined gender.

| Gender | Age group | Wave 1 | Wave 2 | Wave 3 | Wave 4 | Wave 5 | Wave 6 | All waves |
|---|---|---|---|---|---|---|---|---|
| All | All | 184 deaths | 250 deaths | 572 deaths | 686 deaths | 122 deaths | 33 deaths | 1,847 deaths |
|  | 0 - 19 | 1.09% | 0.00% | 0.52% | 0.15% | 1.64% | 0.00% | 0.43% |
|  | 20 - 29 | 1.09% | 0.00% | 0.52% | 1.17% | 0.00% | 3.03% | 0.76% |
|  | 30 - 39 | 4.89% | 2.00% | 4.37% | 4.96% | 0.82% | 0.00% | 4.01% |
|  | 40 - 49 | 2.17% | 3.20% | 9.62% | 13.56% | 1.64% | 3.03% | 8.83% |
|  | 50 - 59 | 10.87% | 8.40% | 14.86% | 21.87% | 4.10% | 0.00% | 15.21% |
|  | ≥ 60 | 79.89% | 86.40% | 70.10% | 58.31% | 91.80% | 93.94% | 70.76% |
| Female | All | 81 deaths | 101 deaths | 242 deaths | 262 deaths | 45 deaths | 17 deaths | 748 deaths |
|  | All | 44.02% | 40.40% | 42.31% | 38.19% | 36.89% | 51.52% | 40.50% |
|  | 0 - 19 | 0.54% | 0.00% | 0.35% | 0.00% | 1.64% | 0.00% | 0.27% |
|  | 20 - 29 | 0.54% | 0.00% | 0.17% | 0.58% | 0.00% | 3.03% | 0.38% |
|  | 30 - 39 | 1.09% | 0.40% | 1.75% | 1.75% | 0.00% | 0.00% | 1.35% |
|  | 40 - 49 | 0.54% | 0.40% | 1.75% | 3.50% | 0.82% | 0.00% | 2.00% |
|  | 50 - 59 | 4.35% | 3.20% | 6.64% | 7.29% | 0.82% | 0.00% | 5.68% |
|  | ≥ 60 | 36.96% | 36.40% | 31.64% | 25.07% | 33.61% | 48.48% | 30.81% |
| Male | All | 103 deaths | 149 deaths | 330 deaths | 422 deaths | 77 deaths | 16 deaths | 1,097 deaths |
|  | All | 55.98% | 59.60% | 57.69% | 61.52% | 63.11% | 48.48% | 59.39% |
|  | 0 - 19 | 0.54% | 0.00% | 0.17% | 0.15% | 0.00% | 0.00% | 0.16% |
|  | 20 - 29 | 0.54% | 0.00% | 0.35% | 0.58% | 0.00% | 0.00% | 0.38% |
|  | 30 - 39 | 3.80% | 1.60% | 2.62% | 3.21% | 0.82% | 0.00% | 2.65% |
|  | 40 - 49 | 1.63% | 2.80% | 7.87% | 10.06% | 0.82% | 3.03% | 6.82% |
|  | 50 - 59 | 6.52% | 5.20% | 8.22% | 14.29% | 3.28% | 0.00% | 9.42% |
|  | ≥ 60 | 42.93% | 50.00% | 38.46% | 33.24% | 58.20% | 45.45% | 39.96% |

**Table S2.** Metrics associated with the COVID-19 confirmed cases and the instantaneous reproduction number in Maringá, fice cities of similar size, and Brazil. The table shows information about the estimated 2021 population, number of total cases, number of weekly cases per 100,000 inhabitants, peak number of weekly cases per 100,000 inhabitants, maximum value of the instantaneous reproduction number $\mathscr{R}_{max}$, proportion of days with $\mathscr{R}(t) > 1$, and average number of consecutive days with $\mathscr{R}(t) > 1$ ($\pm$ one standard deviation). The time window used to calculate these metrics spanned from the date of first case to 22 August 2021, when we have available data for all cities and Brazil.

| Place | Population | Total cases | Weekly cases per 100k | Peak weekly cases per 100k | $\mathscr{R}_{max}(t)$ | Proportion of days with $\mathscr{R}(t) > 1$ | Average number of consecutive days with $\mathscr{R}(t) > 1$ |
|---|---|---|---|---|---|---|---|
| Maringá | 436,472 | 62,327 | 192 | 500 | 2.47 | 0.59 | 16.56 $\pm$9.64 |
| Campina Grande | 413,830 | 42,778 | 141 | 488 | 2.23 | 0.52 | 14.59$\pm$13.70 |
| Piracicaba | 410,275 | 64,520 | 179 | 527 | 2.35 | 0.52 | 12.14$\pm$10.78 |
| Rio Branco | 419,452 | 38,206 | 107 | 384 | 2.13 | 0.44 | 11.25$\pm$10.70 |
| Santos | 433,991 | 62,487 | 164 | 431 | 1.65 | 0.55 | 13.14$\pm$12.05 |
| São José do Rio Preto | 469,173 | 96,187 | 229 | 672 | 2.49 | 0.53 | 13.84$\pm$13.55 |
| Brazil | 213,317,639 | 20,595,014 | 125 | 253 | 4.27 | 0.59 | 19.38$\pm$22.46 |

**Table S3.** Metrics associated with the deaths caused by COVID-19 in Maringá, five cities of similar size, and Brazil. The table shows information about the estimated 2021 population, number of total deaths, number of weekly deaths per 100,000 inhabitants, and peak number of weekly deaths per 100,000 inhabitants for Maringá, five cities of similar size, and Brazil. The time window used to calculate these metrics spanned from the date of the first death to 12 June 2022 (when information was available for all cities and Brazil).

| Place | Population | Total deaths | Weekly deaths per 100k | Peak weekly deaths per 100k |
|---|---|---|---|---|
| Maringá | 436,472 | 1,847 | 3.73 | 23.37 |
| Campina Grande | 413,830 | 1,630 | 1.98 | 13.05 |
| Piracicaba | 410,275 | 1,837 | 2.25 | 16.57 |
| Rio Branco | 419,452 | 1,443 | 1.73 | 19.31 |
| Santos | 433,991 | 2,837 | 3.28 | 23.96 |
| São José do Rio Preto | 469,173 | 3,363 | 3.60 | 30.05 |
| Brazil | 213,317,639 | 681,934 | 1.61 | 11.20 |

**Table S4.** Metrics associated with deaths caused by COVID-19 for each age group in Maringá and Brazil. The table shows information, for each age group, about the total deaths, percentage of deaths, percentage of female deaths, percentage of male deaths, deaths per 100,000 inhabitants, female deaths per 100,000 inhabitants, and male deaths per 100,000 inhabitants in Maringá and Brazil. The 2021's demographic structure of each region was estimated by using the 2010's age and gender structures further rescaled by the ratio between the population estimation of 2021 and the population in 2010. The percentage of deaths for female and male population does not sum up to a hundred per cent because the total number of casualities include deaths classified as undefined gender.

| Age group | | 0-19 | 20-29 | 30-39 | 40-49 | 50-59 | ≥ 60 |
|---|---|---|---|---|---|---|---|
| Total deaths | Maringá | 8 | 14 | 74 | 163 | 281 | 1,307 |
| | Brazil | 4,650 | 6,422 | 22,976 | 53,044 | 93,487 | 478,600 |
| Percentage of deaths | Maringá | 0.43% | 0.76% | 4.01% | 8.83% | 15.21% | 70.76% |
| | Brazil | 0.71% | 0.97% | 3.49% | 8.05% | 14.18% | 72.61% |
| Percentage of deaths (female) | Maringá | 0.27% | 0.38% | 1.36% | 2.01% | 5.69% | 30.84% |
| | Brazil | 0.31% | 0.47% | 1.42% | 3.15% | 5.69% | 33.01% |
| Percentage of deaths (male) | Maringá | 0.16% | 0.38% | 2.66% | 6.83% | 9.43% | 40.00% |
| | Brazil | 0.36% | 0.50% | 2.07% | 4.89% | 8.49% | 39.59% |
| Deaths per 100k people | Maringá | 5.58 | 13.83 | 88.94 | 203.73 | 462.14 | 2,016.82 |
| | Brazil | 5.91 | 14.95 | 62.00 | 170.73 | 405.90 | 1,858.61 |
| Deaths per 100k people (female) | Maringá | 8.65 | 16.77 | 70.77 | 105.68 | 387.15 | 1,923.23 |
| | Brazil | 5.91 | 16.18 | 55.18 | 144.86 | 346.22 | 1,701.58 |
| Deaths per 100k people (male) | Maringá | 5.04 | 17.03 | 149.64 | 413.89 | 769.16 | 3,149.65 |
| | Brazil | 6.63 | 17.26 | 84.12 | 240.05 | 573.06 | 2,548.46 |



\end{document}